\documentclass[12pt]{article}
\usepackage{geometry} 
\usepackage{amssymb}
\usepackage{amsmath}
\usepackage{MnSymbol}
\usepackage{hyperref}
\usepackage{braket}
\usepackage{subcaption,tikz}
\usetikzlibrary{arrows,decorations.markings,decorations.pathreplacing}

\newcommand{\be}{\begin{equation}}
\newcommand{\ee}{\end{equation}}
\newcommand{\bea}{\begin{eqnarray}}
\newcommand{\eea}{\end{eqnarray}}

\newcommand{\al}{\alpha}
\renewcommand{\d}{\delta}
\newcommand{\e}{\epsilon}
\newcommand{\G}{\Gamma}

\newcommand{\la}{\lambda}

\newcommand{\om}{\omega}

\newcommand{\C}{\mathbb{C}}

\newcommand{\rr}{\rightarrow}

\newcommand{\Z}{\mathbb{Z}}

\newcommand{\U}{\operatorname{U}}

\newcommand{\lp}{\left(}
\newcommand{\rp}{\right)}
\newcommand{\ls}{\left[}
\newcommand{\rs}{\right]}

\newcommand{\btau}{\bar{\tau}}

\numberwithin{equation}{section}

\begin{document}
	
\begin{center}
	
	{\large\bf Anomalies, Extensions and Orbifolds}
	
	\vspace*{0.2in}
	
	Daniel G. Robbins$^1$, Eric Sharpe$^2$,
	Thomas Vandermeulen$^1$
	
	\vspace*{0.1in}
	
	\begin{tabular}{cc}
		{\begin{tabular}{l}
				$^1$ Department of Physics\\
				University at Albany\\
				Albany, NY 12222 \end{tabular}} &
		{\begin{tabular}{l}
				$^2$ Department of Physics MC 0435\\
				850 West Campus Drive\\
				Virginia Tech\\
				Blacksburg, VA  24061 \end{tabular}}
	\end{tabular}
	
	{\tt dgrobbins@albany.edu},
	{\tt ersharpe@vt.edu},
	{\tt tvandermeulen@albany.edu}
	
\end{center}
\pagenumbering{gobble}

We investigate gauge anomalies in the context of orbifold conformal field theories.  Such anomalies manifest as failures of modular invariance in the constituents of the orbifold partition function.  We review how this irregularity is classified by cohomology and how extending the orbifold group can remove it.  Working with such extensions requires an understanding of the consistent ways in which extending groups can act on the twisted states of the original symmetry, which leads us to a discrete-torsion like choice that exists in orbifolds with trivially-acting subgroups.  We review a general method for constructing such extensions and investigate its application to orbifolds.  Through numerous explicit examples we test the conjecture that consistent extensions should be equivalent to (in general multiple copies of) orbifolds by non-anomalous subgroups.

\begin{flushleft}
June 2021
\end{flushleft}

\newpage

\tableofcontents

\newpage

\section{Introduction}
\label{sec:intro}
\pagenumbering{arabic}

When calculating the torus partition function of an orbifold CFT, one forms \textit{partial traces} which are group-twisted versions of the parent theory partition function.  For $g_1$ and $g_2$ commuting elements of our symmetry group $G$, these objects can be expressed as\footnote{In practice, a path-integral definition of the partial traces is feasible only for non-anomalous theories of free fields.  In section~\ref{sec:tdl} we will sketch a definition of the partial traces using topological defect lines (TDLs) that can be used more generally.}
\be
\label{pt}
Z_{g_1,g_2}(\tau,\btau)=\int\mathcal{D}\varphi_{g_1,g_2}e^{-S_E[\varphi]}
\ee
where the fields appearing in the path integral are subject to $g_1$ and $g_2$-periodic boundary conditions on the torus' homotopy cycles and $S_E$ is the theory's euclideanized action.  The partition function of the orbifold theory is formed from these objects as
\be
\label{orb_pt}
\frac{1}{|G|}\sum_{\substack{g_1,g_2\in G \\ g_1g_2=g_2g_1}}Z_{g_1,g_2}(\tau,\bar{\tau}).
\ee
By considering the action of genus one modular transformations on the boundary conditions in the path integral, one finds that partial traces (\ref{pt}) should transform into each other as \cite{Dijkgraaf:1989hb}
\be
\label{ptmodtrans}
Z_{g_1,g_2}\left(\frac{a\tau+b}{c\tau+d},\frac{a\btau+b}{c\btau+d}\right)=Z_{g_2^{-c}g_1^a,g_2^dg_1^{-b}}(\tau,\btau).
\ee
where 
\begin{equation}
\left(\begin{matrix}a&b\\c&d\end{matrix}\right)\in\text{SL}(2;\Z). 
\end{equation} 
For readability we will omit one or more arguments from $Z$ going forward.

Now, in a given theory there can exist symmetries for which (\ref{ptmodtrans}) will not hold -- such symmetries are said to have a gauge ('t Hooft) anomaly.  For an example we can look to the compact free boson, which has cyclic $\Z_N$ symmetries given by coordinate shifts, by either the circle coordinate or its dual.  For a concrete example, consider a shift of order two.  This generates a $\Z_2$ symmetry.  If this shift is by either the coordinate or its dual, we can construct a well-behaved orbifold.  However, if we use the $\Z_2$ generated by simultaneous shifts of the coordinate and dual coordinate, the symmetry is anomalous.  We would detect this oddity by modular transformations, e.g. for this symmetry we would find
\be
\label{z2violation}
Z_{1,0}(\tau+2)=-Z_{1,0}(\tau),
\ee
in violation of (\ref{ptmodtrans}).  Ultimately this issue would lead to a failure of modular invariance in the orbifold theory.  Note, however, that this is a fairly mild violation, and in fact by repeated application of (\ref{z2violation}) we have $Z_{1,0}(\tau+4)=Z_{1,0}(\tau)$.  So while the action in question is not consistent as an order two symmetry, it may have a consistent interpretation as something of order four.

In pursuit of this idea, we can append a second $\Z_2$ element to our indices, enlarging $G$ to a new group $\Gamma$ whose elements we will write as $(i,j)$.  This allows us to relabel $Z_{1,0}(\tau+2)$ as its own object: $Z_{(0,1),(1,0)}(\tau)$.  By comparison with (\ref{ptmodtrans}) we can determine the group relations in $\Gamma$.  For example, we would find that $(0,1)^2=(1,0)$, so in fact we see that $\Gamma\cong\Z_4$.  Following this line further, we generate sixteen partial traces for the putative $\Gamma$ orbifold, each of which can be identified with one of the original $G$ partial traces (sometimes with an anomalous phase, chosen to be consistent with (\ref{z2violation})):
\begin{align}
\label{z4toz2}
Z_{(0,0),(0,0)}&= Z_{0,0},\\
Z_{(0,0),(0,1)}&= Z_{0,1},\\
Z_{(0,1),(0,0)}&= Z_{1,0},\\
Z_{(0,1),(0,1)}&= Z_{1,1},\\
Z_{(0,0),(1,0)}&= Z_{0,0},\\
Z_{(0,0),(1,1)}&= Z_{0,1},\\
Z_{(0,1),(1,0)}&= -Z_{1,0},\\
Z_{(0,1),(1,1)}&= -Z_{1,1},\\
Z_{(1,0),(0,0)}&= Z_{0,0},\\
Z_{(1,0),(0,1)}&= -Z_{0,1},\\
Z_{(1,1),(0,0)}&= Z_{1,0},\\
Z_{(1,1),(0,1)}&= -Z_{1,1},\\
Z_{(1,0),(1,0)}&= Z_{0,0},\\
Z_{(1,0),(1,1)}&= -Z_{0,1},\\
Z_{(1,1),(1,0)}&= -Z_{1,0},\\\label{z4toz2last}
Z_{(1,1),(1,1)}&= Z_{1,1}.
\end{align}
Now we can form the orbifold by $\Gamma$:
\be
\frac{1}{4}\sum_{i,j,k,l=0}^1Z_{(i,j),(k,l)}(\tau)=Z_{0,0}(\tau).
\ee
We see that the anomalous phases have caused everything besides the parent theory partition function to cancel out!  In retrospect, this is exactly what we should have expected.  The claim was that we were going to take an order two symmetry whose orbifold was inconsistent and extend it to an order four symmetry with a consistent orbifold.  But clearly we're not inventing any new symmetries for the theory here, only using what's already available (after all, the $\Gamma$ partial traces were simply $G$ partial traces decorated with some phases).  So the only consistent symmetry available to us is the trivial one, and the only consistent orbifold we could have found was the trivial one (i.e. the parent theory).

This example demonstrates the notion of using group extensions to `cure' orbifolds by anomalous actions.  Of course we would like to apply this idea more generally -- given an anomalous group $G$, can we always construct an extension $\Gamma$ of $G$, the orbifold by which is consistent?  If so, what form does the $\Gamma$ orbifold take, and what choices are required to define it?  The purpose of this paper is to motivate conjectural answers to these questions through explicit constructions and examples.  We begin in section \ref{sec:actions} with a systematic investigation of how the action of the extending group $K$ can be modified in the $G$-twisted sectors.  When $K$ acts trivially on the parent theory this leads us to a discrete torsion-like choice special to orbifolds with non-effective subgroups.  Section \ref{sec:tdl} returns to the subject of anomalies, employing the technology of topological defect lines (TDLs) to derive some basic relations satisfied by anomalous partial traces.  In particular we will recover the well-known fact that for a theory with a symmetry given by a group $G$, the possible anomalies in that symmetry are classified by the cohomology group $H^3(G,U(1))$ -- for a physical picture of this, see e.g. section 2 of \cite{Johnson_Freyd_2019}.  In section \ref{sec:construction} we present a group extension due to Tachikawa \cite{Tachikawa_2020} which by construction trivializes any anomaly.  We examine this construction in the context of orbifold theories and present a method for refining it.  With all of this technology established, in section \ref{sec:examples} we tackle a number of examples of anomalous theories and demonstrate extensions that resolve them.  For each we give an explicit computation of the resulting partition function and show that it is well-behaved.  Section \ref{sec:conclusion} provides discussion and interpretation of our results, along with a presentation of questions that remain open about resolving anomalous orbifolds in CFT.

Because the extending group acts trivially on the parent theory (though in general non-trivially on the $G$-twisted states), this story naturally requires an understanding of decomposition, which is the study of non-effective group actions in field theories \cite{decomp1}\cite{decomp2}\cite{decomp3}\cite{decomp4}.  Due to this similarity, we are preparing  companion papers \cite{decomp_qs}\cite{decompanomalies} to this one which explore similar material from the more mathematical viewpoint of decomposition.

\section{Group Actions and Extensions}
\label{sec:actions}

Assume we have a CFT (to which we will refer as the \textit{parent theory}) with a symmetry given by a finite group $G$, and we extend $G$ by a finite abelian group $K$ to some $\Gamma$ as in
\be
1 \to K \to \Gamma \to G \to 1.
\ee
For simplicity at the moment we will assume that $K$ is central in $\Gamma$.  Additionally, we assume that the symmetries discussed in this section are non-anomalous -- we will return to the topic of anomalies in the following section.

In order to construct the orbifold by $\Gamma$ we must know how the symmetry $K$ acts on the parent theory, and there are two cases we would like to distinguish here: $K$ may be a genuine symmetry or it may have trivial action on the parent theory.  Either way we have some action of $K$ in mind.  In the context of forming an orbifold by $\Gamma$, however, the action of $K$ on the parent theory does not tell us everything -- there are in general multiple consistent ways in which $K$ could act on the $G$-twisted states (even if its action on the parent theory is trivial).

Specifically, since as a set $\Gamma$ can be written as $K \times G$, we will write a generic element of $\Gamma$ as $(k,g)$ (or sometimes $\gamma$ for brevity).  The partial traces of the $\Gamma$ orbifold are then $Z_{(k_1,g_1),(k_2,g_2)}$ (assuming the two elements of $\Gamma$ commute).  We have in the back of our mind a picture where such a partial trace can be expressed as a sum over states in the $(k_1,g_1)$-twisted Hilbert space $\mathcal{H}_{(k_1,g_1)}$ with a representation $\rho_{(k_1,g_1)}(k_2,g_2)$ of the element $(k_2,g_2)$ inserted acting on those states.

 Heuristically, what should the building blocks of such a representation $\rho$ be?  We have elements of the form $\rho_{(0,g_1)}(0,g_2)$ which should describe how insertions of $G$ act on the $G$-twisted states.  This is fine -- we began by examining the $G$ orbifold before extending to $\Gamma$, so we should know how $G$ acts on its own twisted states.\footnote{This is more imprecise than it might seem at first glance, as the $(g\in G)$-twisted states are not in general the same as the $((0,g)\in\Gamma)$-twisted states.  But again, this argument is purely motivational and should be accepted as a rough sketch.}  Similarly for $\rho_{(k_1,1)}(k_2,1)$, since we know how $K$ acts on the parent theory we could consider the $K$ orbifold on its own and learn how $K$ acts on its own twisted states.
 
 When we consider $\rho_{(0,g_1)}(k_2,1)$, however, there is room for ambiguity.  The easiest possibility is to let $K$ act on the $G$-twisted states just as on the rest of the parent theory.  We will take this as the baseline definition for $\rho_{(0,g_1)}(k_2,1)$ which corresponds to the partial trace
 \be
 \label{gkg}
 Z_{(0,g_1),(k_2,1)}
 \ee
However, nothing is preventing us from modifying the action of $K$ on these states.  We expect that any correction factor $B\in\C$ should be a homomorphism in the chosen element $k_2$ (e.g. if $k_2$ received a certain correction, $k_2^2$ should receive the square of that correction) and depend on the $G$ projection $g_1$ of the twisted sector in which it acts (as we have the requirement that the modification is trivial for $g_1=1$).  So (\ref{gkg}) can plausibly be modified to
\be
\label{gkgmod}
B(k_2,g_1)Z_{(0,g_1),(k_2,1)}.
\ee
So far we have stipulated that $B(k,1)=B(0,g)=1$: the former because the action of $K$ on the parent theory receives no modifications, the latter from the fact that $B$ is a homomorphism in its $K$ argument (and quite reasonably -- we do not wish to modify the $K$ action when no elements of $K$ are acting).  Of course, (\ref{gkg}) is not the most general partial trace.  By modular invariance we should expect a general modification of the form
\be
\label{kgkg}
\frac{B(k_2,g_1)}{B(k_1,g_2)}Z_{(k_1,g_1),(k_2,g_2)}.
\ee

Will any $B$ of this form lead to consistent results?  We should worry that $B$ could violate modular invariance on the torus.  Recall that the partial traces of an orbifold form orbits under SL$(2;\Z)$, which are the sums of all partial traces which can be related by modular transformations.  Most often it will be sufficient and more convenient to work at the level of orbits (versus individual partial traces) when manipulating partition functions, so we will introduce notation to facilitate this.  We denote the orbit of the partial trace $Z_{g,h}$ by $\langle Z_{g,h}\rangle$.  The simplest example of this is a $\Z_2$ orbifold, in which the partition function is the sum of four partial traces which form two orbits:
\be
Z_{\Z_2\text{ orbifold}}=\frac{1}{2}[Z_{0,0}+Z_{0,1}+Z_{1,0}+Z_{1,1}]=\frac{1}{2}[\langle Z_{0,0}\rangle+\langle Z_{0,1}\rangle].
\ee
The orbit with three members could be written as $\langle Z_{0,1}\rangle$, $\langle Z_{1,0}\rangle$ or $\langle Z_{1,1}\rangle$ -- any member suffices to represent the orbit.

By construction each orbit is a modular invariant quantity (the action (\ref{ptmodtrans}) of the modular group is simply to permute the orbit's members), and the partition function, by virtue of being a linear combination of orbits, is itself modular invariant.  Therefore the only way to guarantee modular invariance in general is to demand that $B$ assign the same phase to an entire modular orbit.  This potentially changes \textit{which} linear combination of orbits forms our partition function, but does not change the fact that it is a linear combination of modular invariants.  Note that with this requirement, the phase of any orbit which includes untwisted sector partial traces (i.e. partial traces of the form $Z_{1,g}$) must be trivial by the normalization conditions we've laid out for $B$.  This means that only modular orbits which are disconnected from the untwisted sector could receive such modification.

Still, we are well aware that we cannot simply take arbitrary linear combinations of modular invariant quantities and expect that the result has a consistent interpretation as the partition function of a CFT.  Luckily, the conditions we've laid out so far will constrain the coefficients that can possibly appear, and will in fact fix these phases to values which give consistent CFTs.  Below we apply this setup to a specific example.

\subsection{Example: Effective $1\to\Z_2\to\Z_2\times\Z_2\to\Z_2\to 1$}

We begin with a $\Z_2$ symmetry and extend it by another $\Z_2$ to form a $\Z_2\times\Z_2$ symmetry.  The question is -- can we modify the action of the extending $\Z_2$ in the manner described above?  We could guess at $B$ that satisfy the conditions laid out so far (and in fact from the requirement that $B$ is a homomorphism in $K$ we would immediately arrive at the only non-trivial possbility), but instead we'll use our criteria to derive the answer systematically.

Writing $\Z_2\times\Z_2$ as $\{1,a,b,c\}$, its orbifold partition function contains five orbits:
\begin{align}
\langle Z_{1,1}\rangle&=Z_{1,1},\\
\langle Z_{1,a}\rangle&=Z_{1,a}+Z_{a,1}+Z_{a,a},\\
\langle Z_{1,b}\rangle&=Z_{1,b}+Z_{b,1}+Z_{b,b},\\
\langle Z_{1,c}\rangle&=Z_{1,c}+Z_{c,1}+Z_{c,c},\\
\langle Z_{a,b}\rangle&=Z_{a,b}+Z_{a,c}+Z_{b,a}+Z_{b,c}+Z_{c,a}+Z_{c,b}.
\end{align}
As previously mentioned, our normalization conditions on $B$ along with the requirement (from modular invariance) that we assign consistent phases within orbits implies that orbits that include an untwisted sector partial trace receive no modification.  In fact this is a general feature in orbifold theories, which we can see from expanding partial traces in a $q$-series which is interpreted as a sum over states where the exponents are the weights of the states and the coefficients are their multiplicities.  In the orbifold theory the vacuum state should be the unique weight zero state and should therefore appear with multiplicity one.  The only $q$ expansions with zeroth order terms will be those in the untwisted sector, and in a $G$ orbifold there are $|G|$ such partial traces.  Since the orbifold partition function (\ref{orb_pt}) has a normalization factor of $|G|^{-1}$, all of the untwisted sector partial traces must enter the partition function with unit phase to preserve the vacuum's unit coefficient.  Modular invariance then fixes the phases of the rest of the partial traces in their orbits.  The upshot of this line of reasoning is that only orbits that are disconnected from the untwisted sector have the potential to receive modifications.

In the case of $\Z_2\times\Z_2$ above, only $Z_{a,b}$ is disconnected from the untwisted sector, so in principle it can experience a modified action from the extending $\Z_2$ and enter the partition function with arbitrary coefficient.  The full partition function is therefore
\be
\frac{1}{4}[\langle Z_{1,1}\rangle+\langle Z_{1,a}\rangle+\langle Z_{1,b}\rangle+\langle Z_{1,c}\rangle+\alpha\langle Z_{a,b}\rangle]
\ee
where we have yet to fix $\alpha$.  Using additive notation for the elements of $K$ and $G$, the only potentially non-trivial value of $B$ is $B(1,1)$.  Using (\ref{kgkg}), if the orbit $\langle Z_{a,b}\rangle$ were to receive a modification, it would become (here we'll assume that $c$ generates $K$):
\begin{multline}
B(1,1)Z_{a,b}+B(1,1)Z_{a,c}+B^{-1}(1,1)Z_{b,a}\\
+B(1,1)Z_{b,c}+B^{-1}(1,1)Z_{c,a}+B^{-1}(1,1)Z_{c,b},
\end{multline}
from which we see that modular invariance requires $B(1,1)=B^{-1}(1,1)$, so $\alpha=\pm 1$.

We can double check this result by putting the theory on a higher genus surface.  In particular a surface of genus two has four homotopy cycles, so the analog of a partial trace (\ref{pt}) can have four separate periodicity conditions and is therefore indexed by four group elements.  At genus two, a $\Z_2\times\Z_2$ orbifold has six Sp$(4;\Z)$ orbits, which enter the partition function as
\be
\label{z2z2gen2}
\frac{1}{16}[\langle Z_{1,1,1,1}\rangle+\langle Z_{1,1,1,a}\rangle+\langle Z_{1,1,1,b}\rangle+\langle Z_{1,1,1,c}\rangle+\langle Z_{1,1,a,b}\rangle+\beta\langle Z_{1,a,1,b}\rangle].
\ee
Again, the last of these orbits is disconnected from the untwisted sector and can in principle enter with arbitrary coefficient, but the coefficients of the rest are fixed by modular transformations.  Our genus two surface has the ability to degenerate to two connected tori.  Under such a degeneration, the period matrix (and therefore partition function) can be expanded as a series.  The leading term in this expansion of the genus two partition function should be the genus one partition function on each torus.  Specifically, a genus two $\Gamma$ partial trace has a leading term given by
\be
\label{degen}
Z_{(k_1,g_1),(k'_1,g'_1),(k_2,g_2),(k'_2,g'_2)}\to Z_{(k_1,g_1),(k_2,g_2)}Z_{(k'_1,g'_1),(k'_2,g'_2)}.
\ee
Applying this degeneration to the $\Z_2\times\Z_2$ genus two orbits gives (we will use a convention where the ordering of the genus one partial traces determines which torus they're on)
\begin{align}
\langle Z_{1,1,1,1}\rangle\to&\langle Z_{1,1}\rangle\langle Z_{1,1}\rangle,\\
\langle Z_{1,1,1,a}\rangle\to&\langle Z_{1,1}\rangle\langle Z_{1,a}\rangle+\langle Z_{1,a}\rangle\langle Z_{1,1}\rangle+\langle Z_{1,a}\rangle\langle Z_{1,a}\rangle,\\
\langle Z_{1,1,1,b}\rangle\to&\langle Z_{1,1}\rangle\langle Z_{1,b}\rangle+\langle Z_{1,b}\rangle\langle Z_{1,1}\rangle+\langle Z_{1,b}\rangle\langle Z_{1,b}\rangle,\\
\langle Z_{1,1,1,a}\rangle\to&\langle Z_{1,1}\rangle\langle Z_{1,c}\rangle+\langle Z_{1,c}\rangle\langle Z_{1,1}\rangle+\langle Z_{1,c}\rangle\langle Z_{1,c}\rangle,\\
\langle Z_{1,1,a,b}\rangle\to&\langle Z_{1,a}\rangle\langle Z_{1,b}\rangle+\langle Z_{1,b}\rangle\langle Z_{1,a}\rangle+\langle Z_{1,b}\rangle\langle Z_{1,c}\rangle\\\notag
+&\langle Z_{1,c}\rangle\langle Z_{1,b}\rangle+\langle Z_{1,c}\rangle\langle Z_{1,a}\rangle+\langle Z_{1,a}\rangle\langle Z_{1,c}\rangle\\\notag
+&\langle Z_{a,b}\rangle\langle Z_{a,b}\rangle,\\
\langle Z_{1,a,1,b}\rangle\to&\langle Z_{1,1}\rangle\langle Z_{a,b}\rangle+\langle Z_{a,b}\rangle\langle Z_{1,1}\rangle+\langle Z_{1,a}\rangle\langle Z_{a,b}\rangle\\\notag
+&\langle Z_{a,b}\rangle\langle Z_{1,a}\rangle+\langle Z_{1,b}\rangle\langle Z_{a,b}\rangle+\langle Z_{a,b}\rangle\langle Z_{1,b}\rangle\\\notag
+&\langle Z_{1,c}\rangle\langle Z_{a,b}\rangle+\langle Z_{a,b}\rangle\langle Z_{1,c}\rangle.
\end{align}
Applying these relations to (\ref{z2z2gen2}), we ought to find the square of the genus one partition function.  From the degeneration we find $\langle Z_{a,b}\rangle\langle Z_{a,b}\rangle$ with coefficient 1, whereas in the square it would appear with coefficient $\alpha^2$.  From this we learn that $\alpha^2=1$, and then can further fix $\beta=\alpha$.  

Thus, both genus one modular invariance and demanding a consistent decomposition from genus two fix the $\Z_2\times\Z_2$ torus partition function as
\be
\frac{1}{4}[\langle Z_{1,1}\rangle+\langle Z_{1,a}\rangle+\langle Z_{1,b}\rangle+\langle Z_{1,c}\rangle\pm\langle Z_{a,b}\rangle].
\ee
The choice of plus or minus for the disconnected orbit is the discrete torsion in the $\Z_2\times\Z_2$ orbifold, classified by $H^2(\Z_2\times\Z_2,U(1))\cong\Z_2$ \cite{vafa_dt}.  Here we have come by it as a choice for how the extending $\Z_2$ acts on the twisted states of the original $\Z_2$ (note that not all discrete torsion is of this form).

Let us briefly consider what would happen if we took the extension class to be non-trivial, such that the full group were $\Z_4$ instead of $\Z_2\times\Z_2$.  We could run through the same procedure, but we would be finished almost immediately -- a $\Z_4$ orbifold has no disconnected orbits, so from the outset there is no possibility of non-trivial phases -- the relative phases of all of the partial traces are fixed by modular transformations.  This is corroborated by the fact that $H^2(\Z_4,U(1))$ is trivial, meaning this orbifold carries no choice of discrete torsion.  In the following sections we investigate a twist on this case which will prove more interesting.

\subsection{Non-Effective Actions}

Now we consider the special case where the extending group $K$ has trivial action on the parent theory.  Here we will find that some of our assumptions about modular invariance can be loosened, and this will lead to phase choices which were not possible for generic $K$.  Specifically, because $K$ acts trivially on the parent theory, all of the $\Gamma$ partial traces must simply be their corresponding $G$ partial traces, up to the action of $K$ from (\ref{kgkg}):
\be
\label{decomp}
Z_{(k_1,g_1),(k_2,g_2)}=\frac{B(k_2,g_1)}{B(k_1,g_2)}Z_{g_1,g_2}.
\ee
This process of reducing $\Gamma$ partial traces to $G$ partial traces (and their associated orbifold(s) as a whole) is known as \textit{decomposition} \cite{decomp1}.

Again we must demand modular invariance, but some additional structure has appeared.  To see this, consider the $\Gamma$ modular orbit $\langle Z_{(k_1,g_1),(k_2,g_2)}\rangle$.  We can use (\ref{decomp}) to write any partial trace in this orbit as a $G$ partial trace, possibly with a phase.  Making modular transformations on both sides of the equation, one can see that each $\Gamma$ orbit $\langle Z_{(k_1,g_1),(k_2,g_2)}\rangle$ must consist of an integral number of copies of the associated $G$ orbit $\langle Z_{g_1,g_2}\rangle$.

When $K$ acted effectively, the only way to guarantee that modular invariance of the $\Gamma$ orbits held was to demand that we assign a single phase to an entire $\Gamma$ orbit.  Now, because each $\Gamma$ orbit decomposes to a linear combination of $G$ orbits (in the form of an integer multiple of one orbit), we are free to allow for differing phases between the various copies of the $G$ orbit.  To see that this new choice actually arises in practice, we move to an example.

\subsection{Example: Non-Effective $1\to\Z_2\to\Z_4\to\Z_2\to 1$}

Again we tackle the extension of $\Z_2$ to $\Z_4$, which was briefly mentioned above.  When the extending $\Z_2$ acted effectively, there were no choices to make and the entire orbifold partition function followed from genus one modular transformations.

To begin, we write out the three $\Z_4$ modular orbits:
\begin{align}
\langle Z_{0,0}\rangle&=Z_{0,0},\\
\langle Z_{0,1}\rangle&=Z_{0,1}+Z_{0,3}+Z_{1,0}+Z_{1,1}+Z_{1,2}+Z_{1,3}+Z_{2,1}+Z_{2,3}\\\nonumber
&+Z_{3,0}+Z_{3,1}+Z_{3,2}+Z_{3,3},\\
\langle Z_{0,2}\rangle&=Z_{0,2}+Z_{2,0}+Z_{2,2}.
\end{align}
To identify the $\Z_4$ partial traces with the original $\Z_2$ partial traces, we would identify the element $2\in\Z_4$ as trivially-acting, which in this case means taking the partial trace indices mod 2.  We do this, keeping in mind to use (\ref{decomp}) to add in phases.  The $\langle Z_{0,2}\rangle$ orbit is indexed entirely by elements of $K$, so can have no non-trivial phases -- it decomposes to three copies of the parent theory partition function.\footnote{More generally, anything that decomposes to the parent theory partition function has $g_1=g_2=1$ and by the normalization conditions on $B$ can receive no phases.}  The orbit $\langle Z_{0,1}\rangle$, on the other hand, has the potential to receive non-trivial phases.  Its members decompose to $\Z_2$ partial traces as
\begin{align}
Z_{0,1}&\to Z_{0,1},\\
Z_{0,3}&\to Z_{0,1},\\
Z_{1,0}&\to Z_{1,0},\\
Z_{1,1}&\to Z_{1,1},\\
Z_{1,2}&\to B^{-1}(1,1)Z_{1,0},\\
Z_{1,3}&\to B^{-1}(1,1)Z_{1,1},\\
Z_{2,1}&\to B(1,1)Z_{0,1},\\
Z_{2,3}&\to B(1,1)Z_{0,1},\\
Z_{3,0}&\to Z_{1,0},\\
Z_{3,1}&\to B(1,1)Z_{1,1},\\
Z_{3,2}&\to B^{-1}(1,1)Z_{1,0},\\
Z_{3,3}&\to B(1,1)B^{-1}(1,1)Z_{1,1}=Z_{1,1}.
\end{align}
We see that in order to maintain modular invariance we must have $B^{-1}(1,1)=B(1,1)$.  The $\Z_4$ modular orbits then decompose as
\begin{align}
\langle Z_{0,0}\rangle&\to\langle Z_{0,0}\rangle,\\
\langle Z_{0,1}\rangle&\to\alpha\langle Z_{0,1}\rangle,\\
\langle Z_{0,2}\rangle&\to 3\langle Z_{0,0}\rangle,
\end{align}
where we will call $\alpha = 2(1+B(1,1))$ a \textit{coefficient of decomposition}.  Since we have learned that $B(1,1)=\pm 1$, we know that $\alpha$ can be 0 or 4.

Since this result is a little unusual (we don't often see choices that aren't discrete torsion in orbifolds), we once again look to genus two for confirmation.  Now we have two tools with which we can operate on the genus two partition function -- we could degenerate its partial traces to genus one partial traces, as we did before, or we could use decomposition to turn the genus two $\Z_4$ partial traces into genus two $\Z_2$ partial traces.  If we apply both operations, we will end up with a product of linear combinations of $\Z_2$ partial traces, one on each post-degeneration torus.  It would be odd if applying these operations in different orders led to different results -- whichever route we take, we should find a unique genus one partition function.  For consistency, we will demand that the operations of degeneration and decomposition commute.\footnote{When $K$ acts trivially and the extended group $\Gamma$ is non-abelian, subtleties appear in the degeneration process due to the presence of multiple weight-zero operators.  We give an example of such an extension and discuss this phenomenon further in appendix \ref{appendix:nonabelian}.}

In the present example, the genus two $\Z_4$ partition function takes the form (in terms of its Sp$(4;\Z)$ orbits)
\be
\label{z4gen2}
\frac{1}{16}[\langle Z_{0,0,0,0}\rangle+\langle Z_{0,0,0,1}\rangle+\langle Z_{0,0,0,2}\rangle].
\ee
Under degeneration these orbits behave as
\begin{align}
\langle Z_{0,0,0,0}\rangle\to&\langle Z_{0,0}\rangle\langle Z_{0,0}\rangle,\\
\langle Z_{0,0,0,1}\rangle\to&\langle Z_{0,0}\rangle\langle Z_{0,1}\rangle+\langle Z_{0,1}\rangle\langle Z_{0,0}\rangle+\langle Z_{0,1}\rangle\langle Z_{0,1}\rangle\\
+&\langle Z_{0,1}\rangle\langle Z_{0,2}\rangle+\langle Z_{0,2}\rangle\langle Z_{0,1}\rangle,\\
\langle Z_{0,0,0,2}\rangle\to&\langle Z_{0,0}\rangle\langle Z_{0,2}\rangle+\langle Z_{0,2}\rangle\langle Z_{0,0}\rangle+\langle Z_{0,2}\rangle\langle Z_{0,2}\rangle,
\end{align}
and under decomposition\footnote{(\ref{degen}) and (\ref{decomp}) imply that the genus two phases from $B$ are simply the product of the corresponding genus one phases, which is how we've calculated the coefficients of decomposition at genus two.}
\begin{align}
\langle Z_{0,0,0,0}\rangle\to&\langle Z_{0,0,0,0}\rangle,\\
\langle Z_{0,0,0,1}\rangle\to&4\alpha\langle Z_{0,0,0,1}\rangle,\\
\langle Z_{0,0,0,2}\rangle\to&15\langle Z_{0,0,0,0}\rangle.
\end{align}
If we first decompose and then degenerate, (\ref{z4gen2}) becomes
\be
\frac{1}{16}[16\langle Z_{0,0}\rangle\langle Z_{0,0}\rangle+4\alpha(\langle Z_{0,1}\rangle\langle Z_{0,0}\rangle+\langle Z_{0,0}\rangle\langle Z_{0,1}\rangle+\langle Z_{0,1}\rangle\langle Z_{0,1}\rangle)].
\ee
If instead we perform degeneration then decomposition, we find
\be
\frac{1}{16}[16\langle Z_{0,0}\rangle\langle Z_{0,0}\rangle+4\alpha(\langle Z_{0,0}\rangle\langle Z_{0,1}\rangle+\langle Z_{0,1}\rangle\langle Z_{0,0}\rangle)+\alpha^2\langle Z_{0,1}\rangle\langle Z_{0,1}\rangle].
\ee
In order for these to be equal, we see that we must have $\alpha^2=4\alpha$, so $\alpha$ can be 0 or 4, matching the calculation from the torus.

To summarize, we began with a $\Z_2$ symmetry.  We extended it to $\Z_4$ by another $\Z_2$ symmetry which acts trivially on the parent theory.  We have found that there exist two choices for the $\Z_4$ orbifold, depending on how the extending $\Z_2$ acts in the twisted sector of the original.  The resulting $\Z_4$ partition function takes the form
\be
\frac{1}{4}[4\langle Z_{0,0}\rangle+\alpha\langle Z_{0,1}\rangle]
\ee
where $\alpha$ can be 4 or 0, for a respective trivial or non-trivial action on the twisted states.  This choice, which in some ways resembles a choice of discrete torsion, only exists when the extending group acts trivially.  When $\alpha=0$ and $B(1,1)=-1$, this exactly matches our result from section \ref{sec:intro} -- we find that the $\Z_4$ orbifold simply returns the parent theory.  In our heuristic attempt to `cure' the $\Z_2$ orbifold of its anomaly, we were led to a non-trivial modification of $K$'s action on $G$'s twisted sectors.

\subsection{Classification of Quantum Symmetries}
\label{sec:classification}

Since we will be discussing it regularly, we ought to give this choice of $K$'s action on the $G$-twisted states a name -- we call it the \textit{quantum symmetry}\footnote{This has no relation to the concept of a quantum group.} of the orbifold.  We would like to have some way to classify the distinct choices of quantum symmetry, as $H^2(G,U(1))$ does for discrete torsion.  In order to obtain such a description, we start with the case where $K$ is central in $\Gamma$ -- partial results on the general case where $K$ is not assumed to be central or abelian are discussed in appendix \ref{app:noncentral}.

The setup for this section is that we have an extension $1\to K\to\Gamma\to G\to 1$ with $K$ finite abelian and central in $\Gamma$.  The extension class is $c(g_1,g_2)\in H^2(G,K)$, which we take to be normalized ($c(1,g)=c(g,1)=0\in K$).  We write elements of $\Gamma$ as elements $(k,g)$ of $K\times G$, and the group operation is given by
\be
(k_1,g_1)(k_2,g_2)=(k_1+k_2+c(g_1,g_2),g_1g_2).
\ee

As argued earlier in the section, in order to guarantee general modular invariance, $B$ should assign the same phase to each partial trace in a given $\Gamma$ orbit.  Consider the partial trace $Z_{(k_1,g_1),(k_2,g_2)}$.  The possible modifications to the action of $K$ are given by (\ref{kgkg}):
\be
\label{kgkg2}
\frac{B(k_2,g_1)}{B(k_1,g_2)}Z_{(k_1,g_1),(k_2,g_2)}
\ee
where, as before, $B$ is a homomorphism in its $K$ argument and is normalized such that $B(0,g)=B(k,1)=1$.  Consider the modular transformation
\be
Z_{(k_1,g_1),(k_2,g_2)}(\tau-1)=Z_{(k_1,g_1),(k_1,g_1)(k_2,g_2)}(\tau)=Z_{(k_1,g_1),(k_1+k_2+c(g_1,g_2),g_1g_2)}(\tau).
\ee
Performing this transformation on (\ref{kgkg2}) yields
\be
\frac{B(k_2,g_1)}{B(k_1,g_2)}Z_{(k_1,g_1),(k_1+k_2+c(g_1,g_2),g_1g_2)}.
\ee
If, however, we first made the modular transformation, and then used (\ref{kgkg2}) to assign phases, we would find
\be
\frac{B(k_1+k_2+c(g_1,g_2),g_1)}{B(k_1,g_1g_2)}Z_{(k_1,g_1),(k_1+k_2+c(g_1,g_2),g_1g_2)}.
\ee
The phases we assign should be unambiguous, so these two operations should commute.  Therefore we require that
\be
\label{epsilon_consistency}
\frac{B(k_2,g_1)}{B(k_1,g_2)}=\frac{B(k_1+k_2+c(g_1,g_2),g_1)}{B(k_1,g_1g_2)}.
\ee
Now we need to keep in mind that the partial trace we began with, $Z_{(k_1,g_1),(k_2,g_2)}$, is only defined when the two elements of $\Gamma$ that index it commute.  Specifically, this means that the group elements appearing in (\ref{epsilon_consistency}) must satisfy
\be
c(g_1,g_2)=c(g_2,g_1)\hspace{.5cm}\text{and}\hspace{.5cm}g_1g_2=g_2g_1.
\ee
Rearranging (\ref{epsilon_consistency}) (using the fact that $B$ is a homomorphism in $K$), we get
\be
\label{qsmodinv}
B(k_1,g_1g_2)=B(k_1,g_1)B(k_1,g_2)B(c(g_1,g_2),g_1)
\ee
Focusing for the moment on split extensions (i.e. ones in which the extension class is trivial), we see that modular invariance requires $B$ to be a homomorphism in $G$:
\be
\label{g_homomorphism}
B(k,g_1g_2)=B(k,g_1)B(k,g_2)
\ee
This means that distinct choices of $B$ live in $\text{Hom}(G,\hat{K})$, where $\hat{K}=\text{Hom}(K,U(1))$ is the Pontryagin dual of $K$.

When the extension is not split, however, a quantum symmetry given by an element of $\text{Hom}(G,\hat{K})$ now has a potential obstruction to modular invariance given by $B(c(g_1,g_2),g_1)$.  We have seen above, in the examples where $\Gamma$ was $\Z_2\times\Z_2$ and $\Z_4$, that some quantum symmetries are equivalent to turning on discrete torsion in $\Gamma$ while some are not.  We can check when this will happen by asking when the insertion from our quantum symmetry defines a two-cocycle on $\Gamma$.  We define
\be
\om(\gamma_1,\gamma_2)=\om((k_1,g_1),(k_2,g_2))=B(k_2,g_1).
\ee
Then,
\be
\label{dt_obstruction}
d\om(\gamma_1,\gamma_2,\gamma_3)=\frac{\om(\gamma_2,\gamma_3)\om(\gamma_1,\gamma_2\gamma_3)}{\om(\gamma_1,\gamma_2)\om(\gamma_1\gamma_2,\gamma_3)}=B(c(g_2,g_3),g_1).
\ee
So the obstruction to the quantum symmetry matching a choice of discrete torsion is $B(c(g_2,g_3),g_1)$; we see from (\ref{qsmodinv}) that an insertion of $B\in\text{Hom}(G,\hat{K})$ where this obstruction vanishes is compatible with modular invariance in $\Gamma$.

The case of most interest to us is the one in which $K$ acts trivially on the parent theory.  Because we now have the option of only requiring modular invariance of the $G$ orbits post-decomposition (as opposed to the full $\Gamma$ orbits), the corresponding modular invariance equations involve constrained sums over elements of $k$ and solving for a general form of $B$ becomes more difficult.  At this point we will switch to conjecture and guess, based on the conditions derived above, that distinct quantum symmetries are classified by $H^1(G,\hat{K})$ (more generally $H^1(G,H^1(K,U(1)))$ for $K$ non-abelian).  In support of this claim, in section \ref{sec:examples} we will check that the choices of quantum symmetry in the examples we examine are consistent with $H^1(G,\hat{K})$.  The most straightforward way to extend the analysis of this section would be to allow $K$ not to be central in $\Gamma$, which means we would have a non-trivial action of $G$ on $K$.  Appendix \ref{app:noncentral} contains the beginnings of such a treatment and, while the results there are not sufficient to cover the most general cases, they are satisfactory for all of the non-central examples we present in section \ref{sec:examples}.

\section{Anomaly Basics}
\label{sec:tdl}

\begin{figure}[h]
	\centering
	\begin{tikzpicture}
	\draw[very thick,->] (-1,2) -- (-0.5,1);
	\draw[very thick] (-0.5,1) -- (0,0);
	\draw[very thick,->] (-1,-2) -- (-0.5,-1);
	\draw[very thick] (-0.5,-1) -- (0,0);
	\draw[very thick,->] (0,0) -- (1,0);
	\draw[very thick] (1,0) -- (2,0);
	\node at (-1,2.3) {$g_1$};
	\node at (-1,-2.3) {$g_2$};
	\node at (1,0.3) {$g_1g_2$};
	\end{tikzpicture}
	\caption{A three-way junction.}
	\label{fig:3way}
\end{figure}
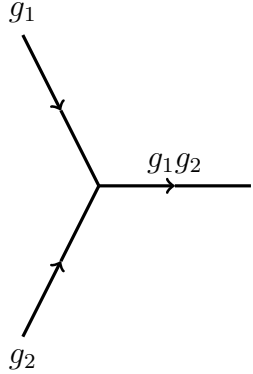

To see the effect of anomalies on the transformations of partial traces, we represent them in terms of topological defect lines (TDLs).  To each line we associate a group element and an orientation specified by an arrow.  Two lines representing the elements $g_1$ and $g_2$ can be joined at a junction to form a line representing $g_1g_2$, as in figure \ref{fig:3way}.

\begin{figure}[h]
	\begin{subfigure}{0.4\textwidth}
		\centering
		\begin{tikzpicture}[scale=1]
		\draw[very thick] (-2,-2) -- (-1.5,-1);
		\draw[very thick,<-] (-1.5,-1) -- (-1,0);
		\draw[very thick] (-2,2) -- (-1.5,1);
		\draw[very thick,<-] (-1.5,1) -- (-1,0);
		\draw[very thick] (-1,0) -- (0,0);
		\draw[very thick,<-] (0,0) -- (1,0);
		\draw[very thick] (2,2) -- (1.5,1);
		\draw[very thick,<-] (1.5,1) -- (1,0);
		\draw[very thick] (2,-2) -- (1.5,-1);
		\draw[very thick,<-] (1.5,-1) -- (1,0);
		\draw[thick,red] (-1,-0.1) -- (-0.8,0.1);
		\draw[thick,red] (-1,0.1) -- (-0.8,-0.1);
		\draw[thick,red] (0.93,0.1) -- (1.17,0.1);
		\draw[thick,red] (1.05,-0.02) -- (1.05,0.22);
		\node at (-2,2.3) {$g_1$};
		\node at (-2,-2.3) {$g_2$};
		\node at (2,-2.3) {$g_3$};
		\node at (2,2.3) {$g_4$};
		\node at (0,0.3) {$g_1g_2$};
		\node at (3,0) {$=$};
		\end{tikzpicture}
	\end{subfigure}
	\begin{subfigure}{0.6\textwidth}
		\centering
		\begin{tikzpicture}[scale=1]
		\draw[very thick] (-2,-2) -- (-1,-1.5);
		\draw[very thick,<-] (-1,-1.5) -- (0,-1);
		\draw[very thick] (-2,2) -- (-1,1.5);
		\draw[very thick,<-] (-1,1.5) -- (0,1);
		\draw[very thick] (2,-2) -- (1,-1.5);
		\draw[very thick,<-] (1,-1.5) -- (0,-1);
		\draw[very thick] (2,2) -- (1,1.5);
		\draw[very thick,<-] (1,1.5) -- (0,1);
		\draw[very thick] (0,-1) -- (0,0);
		\draw[very thick,<-] (0,0) -- (0,1);
		\draw[thick,red] (-0.1,-1) -- (0.1,-0.8);
		\draw[thick,red] (-0.1,-0.8) -- (0.1,-1);
		\draw[thick,red] (0.1,0.93) -- (0.1,1.17);
		\draw[thick,red] (-0.02,1.05) -- (0.22,1.05);
		\node at (-2,2.3) {$g_1$};
		\node at (-2,-2.3) {$g_2$};
		\node at (2,-2.3) {$g_3$};
		\node at (2,2.3) {$g_4$};
		\node at (0.5,0) {$g_2g_3$};
		\node at (4,0) {$\times\text{ }\omega(g_1,g_2,g_3)$};
		\end{tikzpicture}
	\end{subfigure}
	\caption{4-way junctions can be resolved in two distinct but related ways.}
	\label{fig:JunctionResolution}
\end{figure}
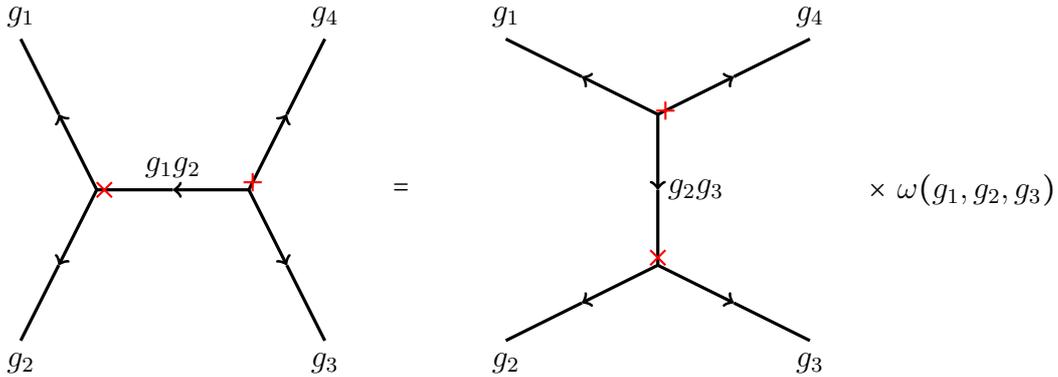

Anomalies appear in this picture when we consider the intersection of four lines associated to $g_1g_2g_3g_4=1$.  There are two ways we could resolve this interaction through trivalent junctions, pictured in figure \ref{fig:JunctionResolution}\footnote{We also should specify an ordering for each trivalent junction that appears.  The convention that we follow is that outgoing lines are ordered clockwise, and in each case the final line in the order is marked with a red x.}.  They turn out to be related by a 3-cocycle $\omega(g_1,g_2,g_3)$ \cite{Chang_2019}.

\begin{figure}
	\begin{subfigure}{\textwidth}
	\centering
	\begin{tikzpicture}
	\draw[thin] (0,0)--(5,0);
	\draw[thin] (5,0)--(5,5);
	\draw[thin] (0,0)--(0,5);
	\draw[thin] (0,5)--(5,5);
	\draw[very thick,->] (2.5,0)--(2,1.25);
	\draw[very thick] (2,1.25)--(1.5,2.5);
	\draw[very thick,->] (3.5,2.5)--(3,3.75);
	\draw[very thick] (3,3.75)--(2.5,5);
	\draw[very thick,->] (5,2.5)--(4.25,2.5);
	\draw[very thick] (4.25,2.5)--(3.5,2.5);
	\draw[very thick,->] (3.5,2.5)--(2.5,2.5);
	\draw[very thick] (2.5,2.5)--(1.5,2.5);
	\draw[very thick,->] (1.5,2.5)--(0.75,2.5);
	\draw[very thick] (0.75,2.5)--(0,2.5);
	\draw[thick,red] (1.2,2.6)--(1.4,2.4);
	\draw[thick,red] (1.2,2.4)--(1.4,2.6);
	\draw[thick,red] (3.2,2.6)--(3.4,2.4);
	\draw[thick,red] (3.2,2.4)--(3.4,2.6);
	\draw[thick,red] (3.2,2.4)--(3.4,2.6);
	\draw[thin,dashed] (0,0)--(-5,5);
	\draw[thin,dashed] (5,0)--(0,5);
	\draw[thin,dashed] (0,5)--(-5,5);
	\draw[thin,dotted] (0,2.5)--(-2.5,2.5);
	\draw[thin,dotted] (-1.5,2.5)--(-2.5,5);
	\node at (3.3,4) {$g$};
	\node at (1.7,1) {$g$};
	\node at (0.7,2.8) {$h$};
	\node at (4.3,2.8) {$h$};
	\node at (2.5,2.8) {$g^{-1}h$};
	\node at (-0.2,5.2) {$\tau+1$};
	\node at (-5.2,5.2) {$\tau$};
	\end{tikzpicture}
	\caption{}
	\label{tanom1}
	\end{subfigure}
	\begin{subfigure}{0.5\textwidth}
	\centering
	\begin{tikzpicture}
	\draw[thin] (0,0)--(5,0);
	\draw[thin] (5,0)--(5,5);
	\draw[thin] (0,0)--(0,5);
	\draw[thin] (0,5)--(5,5);
	\draw[very thick,->] (2.5,0)--(3,1.25);
	\draw[very thick] (3,1.25)--(3.5,2.5);
	\draw[very thick,->] (1.5,2.5)--(2,3.75);
	\draw[very thick] (2,3.75)--(2.5,5);
	\draw[very thick,->] (5,2.5)--(4.25,2.5);
	\draw[very thick] (4.25,2.5)--(3.5,2.5);
	\draw[very thick,->] (3.5,2.5)--(2.5,2.5);
	\draw[very thick] (2.5,2.5)--(1.5,2.5);
	\draw[very thick,->] (1.5,2.5)--(0.75,2.5);
	\draw[very thick] (0.75,2.5)--(0,2.5);
	\draw[thick,red] (1.2,2.6)--(1.4,2.4);
	\draw[thick,red] (1.2,2.4)--(1.4,2.6);
	\draw[thick,red] (3.2,2.6)--(3.4,2.4);
	\draw[thick,red] (3.2,2.4)--(3.4,2.6);
	\node at (1.7,4) {$g$};
	\node at (3.3,1) {$g$};
	\node at (0.7,2.8) {$g^{-1}h$};
	\node at (4.3,2.8) {$g^{-1}h$};
	\node at (2.5,2.8) {$h$};
	\node at (-0.2,5.2) {$\tau$};
	\node at (2.5,-0.75) {};
	\end{tikzpicture}
	\caption{}
	\label{tanom2}
	\end{subfigure}
	\begin{subfigure}{0.5\textwidth}
	\centering
	\begin{tikzpicture}
	\draw[thin] (0,0)--(5,0);
	\draw[thin] (5,0)--(5,5);
	\draw[thin] (0,0)--(0,5);
	\draw[thin] (0,5)--(5,5);
	\draw[very thick,->] (2.5,0)--(2,1.25);
	\draw[very thick] (2,1.25)--(1.5,2.5);
	\draw[very thick,->] (3.5,2.5)--(3,3.75);
	\draw[very thick] (3,3.75)--(2.5,5);
	\draw[very thick,->] (5,2.5)--(4.25,2.5);
	\draw[very thick] (4.25,2.5)--(3.5,2.5);
	\draw[very thick,->] (3.5,2.5)--(2.5,2.5);
	\draw[very thick] (2.5,2.5)--(1.5,2.5);
	\draw[very thick,->] (1.5,2.5)--(0.75,2.5);
	\draw[very thick] (0.75,2.5)--(0,2.5);
	\draw[thick,red] (1.2,2.6)--(1.4,2.4);
	\draw[thick,red] (1.2,2.4)--(1.4,2.6);
	\draw[thick,red] (3.2,2.6)--(3.4,2.4);
	\draw[thick,red] (3.2,2.4)--(3.4,2.6);
	\draw[thick,red] (3.2,2.4)--(3.4,2.6);
	\node at (3.3,4) {$g$};
	\node at (1.7,1) {$g$};
	\node at (0.7,2.8) {$g^{-1}h$};
	\node at (4.3,2.8) {$g^{-1}h$};
	\node at (2.5,2.8) {$g^{-2}h$};
	\node at (-0.2,5.2) {$\tau$};
        \node at (2.5,-0.75) {$\times\ \om(g^{-1},gh^{-1},g)$};
	\end{tikzpicture}
	\caption{}
	\label{tanom3}
	\end{subfigure}
\caption{We evaluate the modular $T$ transformation of an anomalous partial trace.}
\label{anomaloust}
\end{figure}
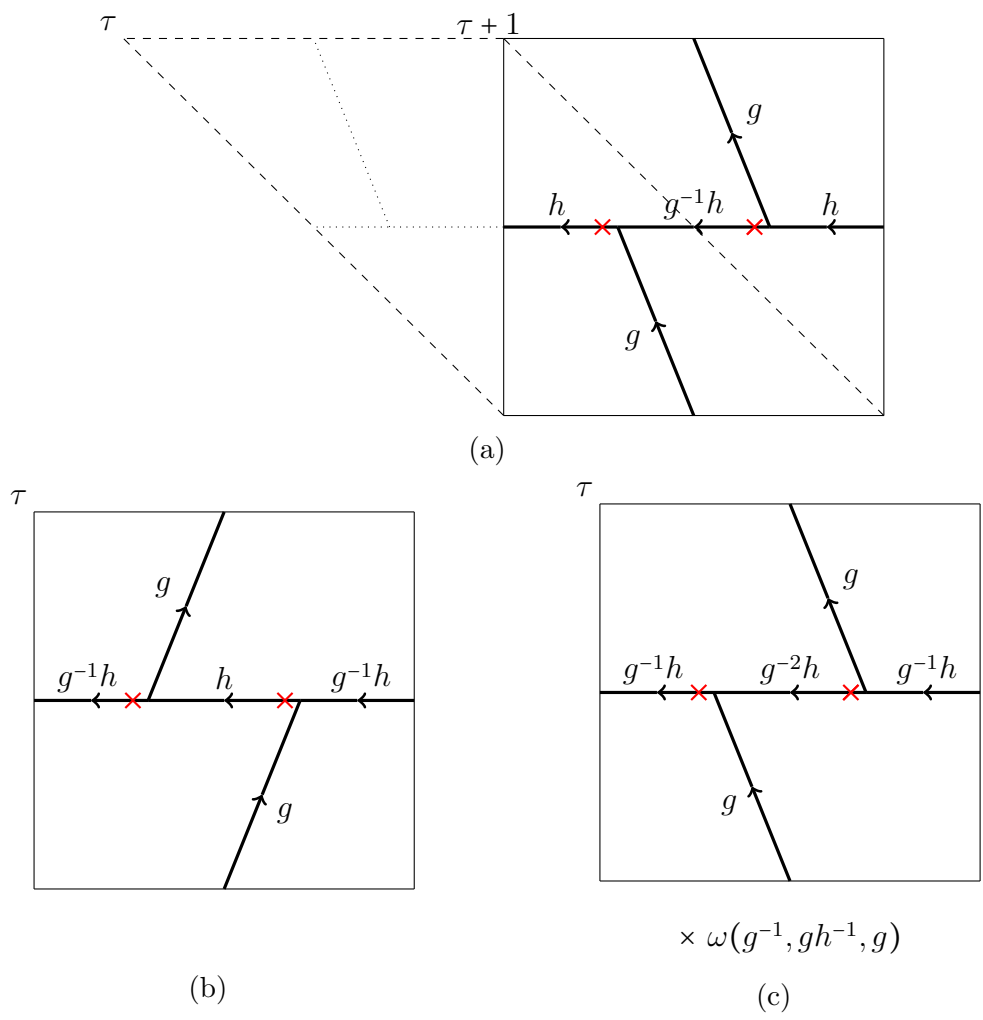

Using this relation, we evaluate the modular $T$ transformation of a partial trace $Z_{g,h}$, the process of which is shown in figure \ref{anomaloust}.  To begin, in \ref{tanom1} we represent the partial trace $Z_{g,h}(\tau+1)$ in terms of TDLs by drawing the group-valued periodicity conditions on the torus' fundamental domain.  This representation is just a choice of convention and it won't necessarily have all of the nice properties of partial traces in the absence of an anomaly.  In particular, its modular transformation is nontrivial and so the modular orbits used to build the orbifold partition functions will need to have extra $\omega$-dependent phases.  For example, to understand the modular $T$ transformation we re-orient the picture to our new fundamental domain indicated by the dashed lines, landing in \ref{tanom2}.  Finally, we use the swap move, picking up a phase $\om(g^{-1},gh^{-1},g)$ and arriving at the picture corresponding to $Z_{g,g^{-1}h}(\tau)$ in figure \ref{tanom3}.  Iterating this process, we find
\be
\label{tanomaly}
Z_{g,h}(\tau+n)=Z_{g,g^{-n}h}(\tau)\prod_{i=1}^n\omega(g^{-1},g^ih^{-1},g)
\ee

In some cases we will assume that the relative phases acquired between different partial traces in these transformations can be absorbed into the definition of the partial trace relative to the TDL picture.  However, we cannot make such accomodations when $n=\text{ord}(g)$.  In that case (\ref{tanomaly}) becomes
\be
\label{tanomalyspec}
Z_{g,h}(\tau+\text{ord}(g))=Z_{g,h}(\tau)\prod_{i=1}^{\text{ord}(g)}\omega(g^{-1},g^{i-1}h^{-1},g)
\ee
and we have a statement purely regarding the modular transformation properties of $Z_{g,h}(\tau)$.  The phase acquired in this way is meaningful in that it cannot be removed via local redefinitions of the traces i.e. it is global.  In fact, we can check that it is something that depends only on the cohomology class of $\omega$ and not on any specific choice of representative.  If we were to shift $\omega$ by an exact quantity $d\lambda$, we would have
\be
\omega(g^{-1},g^{i-1}h^{-1},g)\to \omega(g^{-1},g^{i-1}h^{-1},g)\frac{\lambda(g^{i-1}h^{-1},g)\lambda(g^{-1},g^ih^{-1})}{\lambda(g^{i-2}h^{-1},g)\lambda(g^{-1},g^{i-1}h^{-1})}.
\ee
Under this transformation, the phase in (\ref{tanomalyspec}) becomes
\be
\prod_{i=1}^{\text{ord}(g)}\omega(g^{-1},g^{i-1}h^{-1},g)\prod_{j=1}^{\text{ord}(g)}\frac{\lambda(g^{j-1}h^{-1},g)\lambda(g^{-1},g^jh^{-1})}{\lambda(g^{j-2}h^{-1},g)\lambda(g^{-1},g^{j-1}h^{-1})}.
\ee
Because the product runs over all powers of $g$, the ratio of the various $\lambda$ will be trivial, and as promised our phase depends only on the cohomology class of $\omega\in H^3(G,U(1))$.  This demonstrates that, at least as far as the modular transformations of partial traces are concerned, distinct anomalies are classified by cohomology.

To simplify matters further, one can show that the anomalous phase arising in (\ref{tanomalyspec}) is independent of $h$.  To do this, we examine two specific values of $d\om$:
\be
d\om(g^{-1},g^{i-1},h^{-1},g)=\frac{\om(g^{-1},g^{i-1},h^{-1})\om(g^{-1},g^{i-1}h^{-1},g)\om(g^{i-1},h^{-1},g)}{\om(g^{i-2},h^{-1},g)\om(g^{-1},g^{i-1},h^{-1}g)}
\ee
and
\be
d\om(g^{-1},g^{i-1},g,h)=\frac{\om(g^{-1},g^{i-1},g)\om(g^{-1},g^i,h)\om(g^{i-1},g,h)}{\om(g^{i-2},g,h)\om(g^{-1},g^{i-1},gh)}.
\ee
Noting that many terms will cancel telescopically, we can simplify the product of their ratio to
\be
\prod_{i=2}^{\text{ord}(g)}\frac{d\om(g^{-1},g^{i-1},h^{-1},g)}{d\om(g^{-1},g^{i-1},g,h^{-1})}=\prod_{i=1}^{\text{ord}(g)}\frac{\om(g^{-1},g^{i-1}h^{-1},g)}{\om(g^{-1},g^{i-1},g)}.
\ee
Because $\om$ is a cocycle, the lhs is equal to one and we see that
\be
\prod_{i=1}^{\text{ord}(g)}\om(g^{-1},g^{i-1}h^{-1},g)=\prod_{i=1}^{\text{ord}(g)}\om(g^{-1},g^{i-1},g).
\ee
This version of the result appears in \cite{Gaberdiel:2012gf}.

Having examined modular $T$ transformations, we should not neglect the $S$ transformations.  We can access the phases acquired under $S$ transformations by similar TDL calculations.  As with $T$ transformations, these phases can be absorbed into the definition of the partial traces relative to the TDL picture until we find a relation of a partial trace to itself, which happens after four successive $S$ transformations.  There is the potential for a global anomaly here, but one can check that the phase acquired under $S^4$ is identically trivial.  So in working with anomalous partial traces, we can focus solely on the anomalies that arise in the $T$ transformations of the untwisted sector partial traces.

We now also understand how to determine the anomaly in the extended symmetry $\Gamma$.  Specifically, if we begin with a symmetry $G$ and an $\om\in H^3(G,U(1))$ describing its anomaly, the anomaly in $\Gamma$ will be given by the pullback $\varphi^*\om$ (specifically its cohomology class in $H^3(\Gamma,U(1))$) where $\varphi$ is the homomorphism $\Gamma\to G$ appearing in the extension.

For an explicit example, we look at cyclic $\Z_N$ symmetries.  In \cite{orbits1} it was argued that a $\Z_N$ symmetry with an order $k$ anomaly can be extended to a non-anomalous $\Z_{kN}$ symmetry.  We demonstrate that fact here from the perspective of cohomology.  The relevant anomalies are classified by $H^3(\Z_N,\U(1))\cong\Z_N$ -- we can write the generators of each class as
\be
\label{eq:ZNAnomalyRepresentative}
\omega_m(a,b,c)=\exp{[2\pi ima(\braket{b}+\braket{c}-\braket{b+c})/N^2]}
\ee
where $m\in\{0,\cdots,N-1\}$ determines the class in $H^3(\Z_N,\U(1))$ and $a,b,c$ are integers which label the elements of $\Z_N$; $\braket{z}\equiv z\text{ mod }N$ maps integers into the range $\{0,\cdots,N-1\}$.  Let $k=N/\gcd(N,m)$ (so $k$ is the order of the element $m\in\Z_N$).  There is a natural projection $\varphi$ from $\Z_{kN}$ to $\Z_N$ which is reduction modulo $N$, and $\varphi^\ast\om_m(a,b,c)$ will be given by the expression (\ref{eq:ZNAnomalyRepresentative}), now taking $a,b,c\in\{0,\cdots,kN-1\}$.

We can construct an explicit trivialization of $\varphi^\ast\om_m$ in $H^3(\Z_{kN},\U(1))$.  Define a two-form which takes elements of $\Z_{kN}=\{0,\cdots,kN-1\}$ as arguments:
\be
\lambda(a,b)=\exp{[-2\pi ima\braket{b}/N^2]}.
\ee
We can calculate $d\lambda$ from $d\lambda(a,b,c)=\lambda(b,c)\lambda^{-1}(a+b,c)\lambda(a,b+c)\lambda^{-1}(a,b)$, which yields
\be
d\lambda(a,b,c)=\exp{[2\pi ima(\braket{b}+\braket{c}-\braket{b+c})]/N^2]}=\varphi^\ast\om_m(a,b,c).
\ee
So $\omega_m$ indeed pulls back to something trivial in cohomology.  As a result, the extended $\Z_{kN}$ symmetry has vanishing anomaly.

Finally, we mention that there can exist subgroups of $G$ which display no anomaly, even if $G$ is anomalous.  We call such subgroups non-anomalous.  Let $H$ be a subgroup of $G$ with $\phi:H\to G$ inclusion and $\omega\in H^3(G,U(1))$ giving the anomaly in $G$.  If the pullback $\phi^*\om$ is in the trivial class of $H^3(H,U(1))$, the subgroup $H$ is non-anomalous.  With this definition, it is clear from (\ref{tanomalyspec}) that none of the partial traces indexed purely by elements of $H$ experience anomalous modular transformations.  Note that there always exists at least one non-anomalous subgroup: the trivial one.  Non-anomalous subgroups will be important going forward, as they describe the consistent symmetries of our theory.

In fact, we can now properly state what we mean by resolving or trivializing an anomaly in an orbifold.  We begin with a symmetry $G$ having anomaly $\om\in H^3(G,U(1))$.  We extend $G$ to a larger group $\Gamma$ by $1\to K\to \Gamma\to G\to 1$ where $K$ acts trivially on the parent theory.  We form the orbifold by $\Gamma$ (making any relevant choices, such as discrete torsion and quantum symmetry) and apply decomposition to send all the $\Gamma$ orbits to $G$ orbits.  The general form of the result will be a direct sum of orbifolds by subgroups of $G$.  If all subgroups appearing in this decomposition are non-anomalous, the result is consistent and we say that the extension to $\Gamma$ has resolved the anomaly in the $G$ orbifold.

\section{Tachikawa's Extension}
\label{sec:construction}

In this section we will examine a group extension which for any symmetry $G$ and anomaly $\om$ will guarantee that $\om$ pulls back to the trivial class in $H^3(\Gamma,U(1))$.  (A proof that such an extension always exists originates from the condensed matter literature in \cite{Wang_2018}.)  After presenting the construction we note that it naturally produces an element of $H^1(G,\hat{K})$ which leads to a natural choice of quantum symmetry.  When we examine specific examples of orbifolds by this extended group, we will see that they are in general larger than necessary, i.e. we end up with multiple redundant copies of the cured theory.  We present a method for reducing the result to something minimal, which will be used in subsequent examples.

\subsection{Original construction}
\label{sec:tachikawa}

Let $G$ be a finite group, with $\om\in H^3(G,\U(1))$ its (potential) anomaly.  In \cite{Tachikawa_2020}, Tachikawa gives an explicit construction, which we now review, of a finite group extension
\be
1\longrightarrow A\stackrel{i}{\longrightarrow}\G\stackrel{\varphi}{\longrightarrow} G\longrightarrow 1,
\ee
such that $\varphi^\ast\om$ is exact, i.e.\ there exists $\la\in C^2(\G,\U(1))$ such that $d\la=\varphi^\ast\om$.

Define $A$ as a free $\Z$-module generated by elements $x_{g_1,g_2}$ for $g_1,g_2\in G$, with the additional restriction that $x_{1,g}=x_{g,1}=0$ for all $G$ (where $1$ is the identity element in $G$).  We would like $x_{g,h}$, taken as an element of $C^2(G,A)$, to serve as our extension class -- this requires it to be closed.  The closure condition it must satisfy is
\be
0=g_1\cdot x_{g_2,g_3}+x_{g_1,g_2g_3}-x_{g_1,g_2}-x_{g_1g_2,g_3}.
\ee
We can guarantee this equation holds if we use it to define an action of $G$ on $A$ as
\be
\label{gaction}
g_1\cdot x_{g_2,g_3}=x_{g_1g_2,g_3}-x_{g_1,g_2g_3}+x_{g_1,g_2},
\ee
making $A$ a left $G$-module.

  Explicitly, we define $\G$ as the set $A\times G=\{(m,g)|m\in A,g\in G\}$, and we specify the group multiplication in $\G$ as
\be
(m,g)(n,h):=(m+g\cdot n+x_{g,h},gh).
\ee
This is associative because of the way we defined the group action,
\begin{multline}
((m,g)(n,h))(p,k)=(m+g\cdot n+x_{g,h},gh)(p,k)\\
=(m+g\cdot n+x_{g,h}+gh\cdot p+x_{gh,k},ghk),
\end{multline}
\begin{multline}
(m,g)((n,h)(p,k))=(m,g)(n+h\cdot p+x_{h,k},hk)\\
=(m+g\cdot n+gh\cdot p+g\cdot x_{h,k}+x_{g,hk},ghk),
\end{multline}
and these are equal using the definition (\ref{gaction}) of $g\cdot x_{h,k}$.  The map $\varphi:\G\rr G$ is projection, $\varphi((m,g))=g$, and the map $i:A\rr\G$ is inclusion, $i(m)=(m,1)$.

In order to show that the pullback of $\Gamma$ is trivial, we define
\be
\label{bomegarel}
(b(g))(x_{h,k}):=\om(g,h,k).
\ee
We can extend this by linearity so that for each $g\in G$, $b(g)$ is a homomorphism from $A$ to $\U(1)$.  These maps additionally satisfy a sort of crossed homomorphism condition\footnote{A bit more precisely, if we define $(\hat{b}(g))(m)=(b(g))(g^{-1}\cdot m)$, then $\hat{b}(g)$ is a crossed homomorphism in the usual sense, 
\be
(\hat{b}(gh))(m)=(\hat{b}(h))(g^{-1}\cdot m)(\hat{b}(g))(m)=(g\cdot\hat{b}(h))(m)(\hat{b}(g))(m).
\ee
This means that $\hat{b}$ is coclosed as an element of $C^1(G,\operatorname{Hom}(A,\U(1)))$.  However, we will continue to use the closely related function $b$ in this section, partly to match the convention in~\cite{Tachikawa_2020}.}
\be
\label{eq:bClosedIdentity}
(b(gh))(m)=(b(h))(m)(b(g))(h\cdot m).
\ee
To see that this is true, note that
\begin{multline}
\frac{(b(g))(h\cdot x_{k,\ell})\ (b(h))(x_{k,\ell})}{(b(gh))(x_{k,\ell})}=\frac{(b(g))(x_{hk,\ell}-x_{h,k\ell}+x_{h,k})\ (b(h))(x_{k,\ell})}{(b(gh))(x_{k,\ell})}\\
=\frac{\om(g,hk,\ell)\om(g,h,k\ell)^{-1}\om(g,h,k)\ \om(h,k,\ell)}{\om(gh,k,\ell)}=d\om(g,h,k,\ell)=1,
\end{multline}
since $\om$ is closed.

Finally, in order to check that the $G$ anomaly has been trivialized, we define $\la\in C^2(\G,\U(1))$ by
\be
\la((m,g),(n,h)):=(b(g))(n).
\ee
By explicit computation we have
\begin{multline}
d\la((m,g),(n,h),(p,k))=\frac{\la((n,h),(p,k))\la((m,g),(n+h\cdot p+x_{h,k},hk))}{\la((m+g\cdot n+x_{g,h},gh),(p,k))\la((m,g),(n,h))}\\
=\frac{(b(h))(p)\ (b(g))(n+h\cdot p+x_{h,k})}{(b(gh))(p)\ (b(g))(n)}=\ls\frac{(b(h))(p)\ (b(g))(h\cdot p)}{(b(gh))(p)}\rs (b(g))(x_{h,k})\\
=\om(g,h,k)=\varphi^\ast\om((m,g),(n,h),(p,k)).
\end{multline}
So we find that the pullback of $\omega$ is exact and therefore trivial in cohomology.  As promised, the extended group $\Gamma$ has vanishing anomaly.  This leads us to ask what form the orbifold by this $\Gamma$ takes.

\subsection{Application to Orbifolds}
\label{sec:orbapp}

As discussed in section \ref{sec:actions}, in forming the orbifold by $\Gamma$ there may be multiple consistent choices of quantum symmetry, the action of $A$ on the $G$-twisted states.  We argued that such actions should be classified by $H^1(G,\hat{A})$.  In running through this construction we have in fact come across such a cocycle -- the quantity $b$ defined in (\ref{bomegarel}).  From (\ref{decomp}), then, we should relate the $\Gamma$ partial traces to $G$ partial traces by
\be
\label{bphases}
Z_{(a_1,g_1),(a_2,g_2)}=\frac{(b(g_1))(a_2)}{(b(g_2))(a_1)}Z_{g_1,g_2}.
\ee
Using the definition of $b$, we can rewrite the above in terms of $\om$.  Writing
\be
a_1=\sum_{g,h\in G\backslash\{1\}}n^{(1)}_{g,h}x_{g,h},\qquad a_2=\sum_{g,h\in G\backslash\{1\}}n^{(2)}_{g,h}x_{g,h},
\ee
as general elements of the $\Z$-module $A$, we have
\be
Z_{(a_1,g_1),(a_2,g_2)}=\prod_{\substack{h_1,h_2 \\ k_1, k_2 \\ \in G\backslash\{1\}}}\om(g_1,h_1,k_1)^{n^{(1)}_{h_1,k_1}}\om^{-1}(g_2,h_2,k_2)^{n^{(2)}_{h_2,k_2}}Z_{g_1,g_2}.
\ee
The full $\Gamma$ orbifold partition function is then
\be
\frac{1}{|G||\Z|^{(|G|-1)^2}}\sum_{\substack{g_1,g_2\in G \\ n^{(1)}, n^{(2)}}}\prod_{\substack{h_1,h_2 \\ k_1, k_2 \\ \in G\backslash\{1\}}}\om(g_1,h_1,k_1)^{n^{(1)}_{h_1,k_1}}\om(g_2,h_2,k_2)^{n^{(2)}_{h_2,k_2}}Z_{g_1,g_2},
\ee
where here $n^{(1)}$ and $n^{(2)}$ in the sum are each the collection of $(|G|-1)^2$ exponents to the $\om$ in the product, and each individually runs over all integers.  We can rearrange this expression to cast the group extension as a modification of the original $G$ orbifold:
\be
\label{fullpf}
\frac{1}{|G|}\sum_{g_1,g_2\in G}Z_{g_1,g_2}\left[\frac{1}{|\Z|^{(|G|-1)^2}}\sum_{n^{(1)},n^{(2)}}\prod_{\substack{h_1,h_2 \\ k_1, k_2 \\ \in G\backslash\{1\}}}\om(g_1,h_1,k_1)^{n^{(1)}_{h_1,k_1}}\om(g_2,h_2,k_2)^{n^{(2)}_{h_2,k_2}}\right].
\ee

Let's see what this becomes for specific examples.  

First we have the case that the theory is not actually anomalous, in which all of the $\om$ are trivial.  The products are then all trivial and performing both $n$ sums gives us the partition function (for simplicity in this toy example we assume $\Gamma$ is abelian):
\be
\label{example1}
\frac{|\Z|^{(|G|-1)^2}}{|G|}\sum_{g_1,g_2\in G}Z_{g_1,g_2},
\ee
an infinite direct product of $G$ orbifolds.

Now we look at the simplest anomalous symmetry, a $\Z_2$ with non-zero anomaly.  The products in (\ref{fullpf}) have only a single term, so we can express the full partition function as
\be
\frac{1}{2}\sum_{g_1,g_2\in\Z_2}Z_{g_1,g_2}\left[\frac{1}{|\Z|}\sum_{n_1,n_2=-\infty}^{\infty}\om(g_1,1,1)^{n_1}\om(g_2,1,1)^{n_2}\right].
\ee
As a reminder we have $\om(0,1,1)=1$ and $\om(1,1,1)=-1$.  The four terms in the $g_1,g_2$ summand are then
\begin{align}
&Z_{0,0}\frac{1}{|\Z|}\sum_{n_1,n_2=-\infty}^{\infty} = |\Z|Z_{0,0}\\
&Z_{0,1}\frac{1}{|\Z|}\sum_{n_1,n_2=-\infty}^{\infty}(-1)^{n_1} = 0\\
&Z_{1,0}\frac{1}{|\Z|}\sum_{n_1,n_2=-\infty}^{\infty}(-1)^{n_2} = 0\\
&Z_{1,1}\frac{1}{|\Z|}\sum_{n_1,n_2=-\infty}^{\infty}(-1)^{n_1+n_2} = 0
\end{align}
so the total orbifold partition function, cured of its anomaly, is
\be
\label{example2}
\frac{|\Z|}{2}Z_{0,0}.
\ee

As (\ref{example1}) and (\ref{example2}) show, an orbifold by $\Gamma$ defined in this way is massively overcounted and contains far more copies of the theory than are necessary to cure the anomaly.  Fortunately, Tachikawa gives a prescription for reducing $A$ to a finite group \cite{Tachikawa_2020}; we will sketch its application in the context of orbifolds.  To begin with, define the quantity $\delta$ as the smallest positive integer for which $\omega(g_1,g_2,g_3)^\delta=1$ for all $g_1,g_2,g_3\in G$.  Looking at the quantity in brackets in (\ref{fullpf}), it should be clear that the various $n$ will be $\delta$-periodic.  So we can safely take each $n$ to range from 0 to $\delta-1$.  For compactness of notation, let's define a new quantity
\be
\Lambda_{g_1,g_2}(\om)=\frac{1}{\delta^{(|G|-1)^2}}\sum_{n^{(1)},n^{(2)}\ \mathrm{mod\ }\d}\prod_{\substack{h_1,h_2 \\ k_1, k_2 \\ \in G\backslash\{1\}}}\om(g_1,h_1,k_1)^{n^{(1)}_{h_1,k_1}}\om(g_2,h_2,k_2)^{n^{(2)}_{h_2,k_2}}.
\ee
For finite $\delta$, $\Lambda$ will be finite.  Now we can express the $G$ orbifold partition function, corrected for anomalies, as
\be
\label{reduced_orb_pf}
\frac{1}{|G|}\sum_{g_1,g_2}Z_{g_1,g_2}\Lambda_{g_1,g_2}(\om).
\ee
With this refinement, we should now have $A\cong\Z_{\delta}^{(|G|-1)^2}$.  Let's re-examine the examples from earlier with this new, finite version of $A$.

When our theory is non-anomalous, we can take $\om$ to be trivial.\footnote{Of course trivial anomaly does not imply that $\omega$ is trivial, only that it's in the trivial class in $H^3(G,U(1))$.  The case of a non-trivial $\omega$ giving a trivial anomaly falls under the following section's discussion of representative-dependence.}  Then $\delta=1$ (in this case we take $A$ to be the trivial group) and $\Lambda_{g_1,g_2}(\omega)=1$ across the board, so (\ref{reduced_orb_pf}) simply becomes the usual orbifold partition function.

For the anomalous $\Z_2$ theory, we will have $\delta=2$ which gives $A\cong \Z_2$.  We find $\Lambda_{0,0}=2$ while $\Lambda_{0,1}=\Lambda_{1,0}=\Lambda_{1,1}=0$.  The full partition function (\ref{reduced_orb_pf}) then becomes a single copy of the parent theory partition function $Z_{0,0}$, which agrees with the calculation in section \ref{sec:intro}.

\subsection{Minimal Extensions}

In both of the examples we've examined, the result of Tachikawa's construction with finite $A$ is that we have a single copy of the anomaly-resolved partition function.  We refer to any extension $\Gamma$ the orbifold by which produces a single copy of a CFT (versus a direct sum) as \textit{minimal}, as this is in some sense the smallest extension we could hope for that resolves the anomaly in that way.\footnote{To give a clarifying example, consider an anomalous $\Z_4$ symmetry whose $\Z_2$ subgroup is non-anomalous.  We could produce extensions $1\to\Z_4\to\Z_{16}\to\Z_4\to 1$ and $1\to \Z_2\to \Z_8\to\Z_4\to 1$ whose orbifolds yield one copy of the parent theory and one copy of the orbifold by the non-anomalous $\Z_2$ subgroup, respectively.  While one of these extensions produces a smaller extended group, by our definition they are both minimal since they both lead to a single copy of a CFT.}

Will reducing $A$ to a finite group always yield a minimal extension?  For a counterexample, let's examine a $\Z_3$ symmetry.  We'll write the 3-cocycles as
\be
\begin{matrix}
	\om(1,1,1)=\al, & \om(1,1,2)=\al\beta^{-1}\zeta, & \om(1,2,1)=\al^{-1}\zeta, & \om(1,2,2)=\al^{-1}\beta\zeta, \\ \om(2,1,1)=\beta, & \om(2,1,2)=\al\zeta^2, & \om(2,2,1)=\beta^{-1}\zeta^2, & \om(2,2,2)=\al^{-1}\zeta^2.
\end{matrix}
\ee
where $\zeta^3=1$ determines the cohomology class and $|\alpha|=|\beta|=1$ parameterize the freedom to shift by coboundaries.  We'll start by picking $\alpha=\beta=1$, while $\zeta=\exp{(2\pi i/3)}$ or $\exp{(4\pi i/3)}$ (both non-trivial choices of anomaly will yield the same results).  Then we find $\Lambda_{0,0}=3^4$ and all other $\Lambda$ vanish.  As expected we recover the parent theory partition function ($\Z_3$ can have no non-anomalous subgroups besides the identity), but with a coefficient of $3^3$.  Recall that this method gave an extending group $A\cong\Z_\delta^{(|G|-1)^2}$, which in this case is $\Z_3^4$.  However, from the $\Z_N$ example presented in section \ref{sec:tdl}, one can construct a trivializing extension with $A=\Z_3$, $\Gamma=\Z_9$ (which is in fact minimal).  So our extending group was too large by a factor of $\Z_3^3$, explaining the $3^3$ overcounting.  Therefore the procedure that we have outlined to generate a finite extension will not always produce a minimal one.

What if we don't make the `obvious' choice of $\alpha=\beta=1$?  A simple example of this will be to leave $\beta=1$ but pick $\alpha=-1$.  Now we will be forced to take $\delta=6$ since $\om$ includes an order two component.  Indeed, if we were to try to leave $\delta$ as 3, we would find that the anomalous partial traces no longer drop out of the partition function.  Properly setting $\delta$ to 6 and calculating the various $\Lambda$, we once again find that only $\Lambda_{0,0}$ is non-vanishing, this time having a value of $6^4$.  So the resulting partition function will be $2^43^3Z_{0,0}$.  By picking a cocycle representative that increased our value of $\delta$, we were forced into adding more copies of the theory, and we found ourselves even farther from a minimal extension.

The most extreme example of this would be if we were to take $\alpha$ and/or $\beta$ to have infinite multiplicative order, i.e.\ the corresponding angles were irrational multiples of $2\pi$.  Then we can no longer find any finite value for $\delta$ and we are forced to use the full expression (\ref{fullpf}) to build the orbifold.  This makes sense when we think in terms of a group action.  While choosing a different representative for $\om$'s cohomology class doesn't change the anomaly in question, it does change the phases (\ref{bphases}) coming from the quantum symmetry.  Having $A$ act by irrational phases means that in order to build a sum where all of the anomalous elements cancel out, we are forced into taking an infinite number of copies of our theory.  This representative-dependence is endemic to the construction and will be important to keep in mind going forward.

\subsection{Refined construction}
\label{sec:refinement}

Here we present a method for reducing $A$ further.  In many cases, when $\om$ is a suitably chosen representative (the caveats of the previous section still apply here), this construction will yield a minimal extension.

Define a submodule $\bar{A}\subset A$ (we can begin from either the finite or original infinite version of $A$ here) by
\be
\bar{A}=\left\{m\in A|(b(g))(m)=1,\ \forall g\in G\right\},
\ee
where $b$ was defined in (\ref{bomegarel}).  This is a submodule by the linearity of $b$ and by the fact that for $v\in \bar{A}$ and any $g,h\in G$, we have, using (\ref{eq:bClosedIdentity}),
\be
(b(g))(h\cdot v)=(b(gh))(v)\ (b(h))(v)^{-1}=1,
\ee
so $h\cdot v\in \bar{A}$.

Now define the quotient space $C=A/\bar{A}$, the equivalence classes $[m]$ of $A$ under the relation $m\sim n$ iff $m-n\in \bar{A}$.  Then we define $\widehat{\G}$ as the set $C\times G$ with group multiplication
\be
([m],g)([n],h)=([m+g\cdot n+x_{g,h}],gh),
\ee
which is well-defined since it doesn't depend on the representatives $m$ and $n$, only on their equivalence classes $[m]$ and $[n]$.  Associativity follows from associativity in $\G$.  Again the map $\varphi:\widehat{\G}\rr G$ is just projection.

Then we can define $\la\in C^2(\widehat{\G},\U(1))$ by
\be
\la(([m],g),([n],h))=(b(g))(n).
\ee
This is well-defined because of the definition of $\bar{A}$, which ensures that if we replace $n$ by a different representative $n+v$, $v\in \bar{A}$, the right hand side remains invariant.  The calculation showing that $d\la=\varphi^\ast\om$ proceeds exactly as before, ensuring that we can continue not to worry about anomalies in $\widehat{\Gamma}$.

\section{Examples}
\label{sec:examples}

We now construct, for various choices of symmetry $G$ and anomaly $\om\in H^3(G,U(1))$, trivializing extensions of $G$.  For most examples our method of choice will be the refinement of Tachikawa's procedure laid out above in section \ref{sec:refinement}.  In each case we show how the anomaly would affect the $G$-orbifold's torus partition function, explicitly construct the orbifold partition function of the extended group $\Gamma$, and show how that theory can be expressed in terms of an orbifold (or direct sum of orbifolds) by non-anomalous subgroup(s) of $G$.

\subsection{$G=\Z_2$}

To begin, we return to the example with which we began the paper, the simple case of an anomalous $\Z_2$ symmetry.  In section \ref{sec:intro} we treated this example in a rather ad hoc fashion; here we will be able to check that the methods we have laid out so far reproduce those results.  It is easy to see that, assuming as always that we are using normalized cocycles, the only element of $H^3(\Z_2,U(1))$ that can be non-trivial is $\omega(1,1,1)$ (in additive notation).  The choice, then, of $\omega(1,1,1)=\pm 1$ will characterize non-anomalous vs. anomalous $\Z_2$ symmetries.  In the following subsections we not only rederive $\Z_4$ as a minimal trivializing extension, but showcase additional, qualitatively different trivializing extensions.

\subsubsection{$1\to\Z_2\to\Z_4\to\Z_2\to 1$}

Let's apply the methods of sections \ref{sec:tachikawa} and \ref{sec:refinement} to find a trivializing extension.  In this case $A=\{nx_{1,1}|n\in\Z\}$.  Taking the non-trivial choice of $\omega$, the cocycle $b$ is given by (\ref{bomegarel}) as
\be
(b(0))(nx_{1,1})=1,\qquad (b(1))(nx_{1,1})=(-1)^n.
\ee
It's then easy to see that $\bar{A}=\{2nx_{1,1}|n\in\Z\}$ and $C=\{[0],[x_{1,1}]\}\cong\Z_2$.  Under this identification, $b$ has a single non-trivial value:
\be
(b(1))([x_{1,1}])=-1.
\ee
This will correspond to choosing the non-trivial element in $H^1(\Z_2,\hat{\Z}_2)\cong\Z_2$.

The larger group $\widehat{\G}$ has four elements, and one readily sees that $\widehat{\G}\cong\Z_4$, since
\be
([0],1)^2=([x_{1,1}],0),\quad ([0],1)^3=([x_{1,1}],1),\quad ([0],1)^4=([0],0).
\ee
Thus the anomaly is trivialized by the extension $1\to\Z_2\to\Z_4\to\Z_2\to 1$, matching our previous result.

To reproduce (\ref{z4toz2})-(\ref{z4toz2last}), we use (\ref{bphases}) to relate the $\Z_4$ partial traces to those of the original $\Z_2$ orbifold.  As in section \ref{sec:actions} we will use $\langle Z_{g,h}\rangle$ to denote the sum of all partial traces in the modular orbit of $Z_{g,h}$.  For a $\Z_2$ orbifold we have two orbits:
\be
\langle Z_{0,0}\rangle=Z_{0,0}\hspace{.5cm}\text{and}\hspace{.5cm}\langle Z_{0,1}\rangle=Z_{0,1}+Z_{1,0}+Z_{1,1}.
\ee

A $\Z_4$ orbifold consists of three orbits:
\begin{align}
\langle Z_{0,0}\rangle&=Z_{0,0},\\
\langle Z_{0,1}\rangle&=Z_{0,1}+Z_{0,3}+Z_{1,0}+Z_{1,1}+Z_{1,2}+Z_{1,3}+Z_{2,1}+Z_{2,3}\\\nonumber
&+Z_{3,0}+Z_{3,1}+Z_{3,2}+Z_{3,3},\\
\langle Z_{0,2}\rangle&=Z_{0,2}+Z_{2,0}+Z_{2,2}.
\end{align}
To express elements of $\widehat{\Gamma}$ in additive notation, we would identify
\be
0=([0],0),\quad 1=([0],1),\quad 2=([x_{1,1}],0),\quad 3=([x_{1,1}],1).
\ee
We can now use (\ref{bphases}) to assign phases from the quantum symmetry.  We'll look at the partial trace $Z_{3,2}$ for an example of this:
\be
Z_{3,2}=Z_{([x_{1,1}],1),([x_{1,1}],0)}=\frac{(b(1))([x_{1,1}])}{(b(0))([x_{1,1}])}Z_{1,0}=-Z_{1,0}.
\ee
Following this procedure for each partial trace, we find that the $\Z_4$ orbits relate to the $\Z_2$ orbits by
\begin{align}
\langle Z_{0,0}\rangle&\to \langle Z_{0,0}\rangle,\\
\langle Z_{0,1}\rangle&\to 0,\\
\langle Z_{0,2}\rangle&\to 3\langle Z_{0,0}\rangle.
\end{align}
So in total the $\Z_4$ orbifold reduces to $\frac{1}{4}[\langle Z_{0,0}\rangle+3\langle Z_{0,0}\rangle]=Z_{0,0}$, the parent theory partition function, as expected.  Reobtaining this by now familiar result is a good sanity check that the procedures outlined in section \ref{sec:construction} are consistent.

\subsubsection{$1\to\Z_2\to\Z_2\times\Z_2\to\Z_2\to 1$}
\label{sec:z2anom}

Our next example demonstrates that we can cure the anomalous $\Z_2$ with an anomalous extension: $1\to\Z_2\to\Z_2\times\Z_2\to\Z_2\to 1$.  The representative $\om$ of the non-trivial class in $H^3(\Z_2,U(1))$ will pull back to a non-trivial class in $H^3(\Z_2\times\Z_2,U(1))$, which in the notation of section \ref{z2z2} we can take to be $(\epsilon_a,\epsilon_b,\epsilon_c)=(1,-1,-1)$.  Since the pullback of $\omega$ does not trivialize, the $\Z_2\times\Z_2$ is anomalous.  Regardless, we will follow through computing its orbifold.  Writing the extended group as $\{1,a,b,c\}\cong\Z_2\times\Z_2$ and the original group as $\{1,g\}\cong\Z_2$, the $\Z_2\times\Z_2$ orbifold is
\be
\label{z2z2pf}
\frac{1}{4}[\langle Z_{1,1}\rangle+\langle Z_{1,a}\rangle+\langle Z_{1,b}\rangle+\langle Z_{1,c}\rangle\pm \langle Z_{a,b}\rangle].
\ee
Letting $a$ generate $K$, in terms of $\Gamma/K$ equivalence classes the elements of $G$ are $1=\{1,a\}$, $g=\{b,c\}$.  Under decomposition, the $\Z_2\times\Z_2$ partition function becomes
\be
\frac{1}{2}[2\langle Z_{1,1}\rangle+(1\pm 1)\langle Z_{1,g}\rangle].
\ee
For the trivial choice of discrete torsion, we find two copies of the anomalous $\Z_2$ orbifold.  However, when we take the discrete torsion to be non-trivial, the anomalous orbit drops out of the partition function and we have a single copy of the parent theory, exactly as in the $1\to\Z_2\to\Z_4\to\Z_2\to 1$ trivializing extension.  We can verify that the anomaly in $\Z_2\times\Z_2$ does not disrupt this conclusion.  Writing the $\Z_2\times\Z_2$ orbifold partition function with non-trivial discrete torsion (which can be regarded as a formal construction as this $\Z_2\times\Z_2$ is anomalous) in terms of its partial traces, we have
\begin{multline}
\frac{1}{4}[Z_{1,1}+Z_{1,a}+Z_{a,1}+Z_{a,a}+Z_{1,b}+Z_{b,1}+Z_{b,b}+Z_{1,c}+Z_{c,1}+Z_{c,c}\\
-Z_{a,b}-Z_{b,a}-Z_{b,c}-Z_{c,b}-Z_{a,c}-Z_{c,a}].
\end{multline}
Under the modular transformation $\tau\to\tau+2$, this partition function becomes
\begin{multline}
\frac{1}{4}[Z_{1,1}+Z_{1,a}+Z_{a,1}+Z_{a,a}+Z_{1,b}-Z_{b,1}-Z_{b,b}+Z_{1,c}-Z_{c,1}-Z_{c,c}\\
-Z_{a,b}+Z_{b,a}+Z_{b,c}+Z_{c,b}-Z_{a,c}+Z_{c,a}].
\end{multline}
which, despite its lack of modular invariance, still reduces to $Z_{1,1}$ when expressed in terms of $\Z_2$ partial traces.

So it would seem that the extension to $\Z_2\times\Z_2$ with non-trivial discrete torsion is a minimal trivializing extension\footnote{Recall in section \ref{sec:actions} we derived the discrete torsion in the $\Z_2\times\Z_2$ orbifold as a choice of the extending group's action.  So turning on the $\Z_2$ discrete torsion in this orbifold is in fact equivalent to choosing the non-trivial quantum symmetry in $H^1(\Z_2,\hat{\Z}_2)\cong\Z_2$.}, despite its apparent lack of modular invariance.  This example serves to illustrate two points.  First, it shows that minimal extensions are not unique (we didn't have any particular reason to think that they should be, but this is the first counterexample).  Second, it demonstrates that the pullback of $\omega$ trivializing is not a necessary condition for ending up with a non-anomalous orbifold.

\subsubsection{$1\to\Z_4\to Q_8\to\Z_2\to 1$}
\label{sec:quaternion_extension}

Next we present an example of a non-central, non-split trivializing extension.  We extend our anomalous $\Z_2$ symmetry by $\Z_4$ to the group of quaternions, $Q_8$.  Writing the elements of $Q_8$ as the set $\Z_4\times\Z_2$, we have the group multiplication
\be
(s,p)(q,r)=(s+p\cdot q+2pr,p+r)
\ee
where the non-trivial element of $\Z_2$ acts on $\Z_4$ as $0\to 0$, $1\to 3$, $2\to 2$ and $3\to 1$.

$H^2(Q_8,U(1))$ is trivial, so $Q_8$ orbifolds do not feature a choice of discrete torsion.  However, with the above action of $\Z_2$ on $\Z_4$, we find that $H^1(\Z_2,\hat{\Z}_4)\cong\Z_2$, so we will be able to make a non-trivial choice of quantum symmetry.  Writing the quaternion group in the more traditional fashion\\
$\braket{i,j,k|i^2=j^2=k^2=ijk=-1}$, its modular orbits are
\begin{align}
\langle Z_{1,1}\rangle&=Z_{1,1},\\
\langle Z_{1,i}\rangle&=Z_{1,i}+Z_{1,-i}+Z_{i,1}+Z_{-i,1}+Z_{i,i}+Z_{i,-1}\\\notag
&+Z_{i,-i}+Z_{-i,i}+Z_{-i,-1}+Z_{-i,-i}+Z_{-1,i}+Z_{-1,-i},\\
\langle Z_{1,j}\rangle&=Z_{1,j}+Z_{1,-j}+Z_{j,1}+Z_{-j,1}+Z_{j,j}+Z_{j,-1}\\\notag
&+Z_{j,-j}+Z_{-j,j}+Z_{-j,-1}+Z_{-j,-j}+Z_{-1,j}+Z_{-1,-j},\\
\langle Z_{1,k}\rangle&=Z_{1,k}+Z_{1,-k}+Z_{k,1}+Z_{-k,1}+Z_{k,k}+Z_{k,-1}\\\notag
&+Z_{k,-k}+Z_{-k,k}+Z_{-k,-1}+Z_{-k,-k}+Z_{-1,k}+Z_{-1,-k},\\
\langle Z_{1,-1}\rangle&=Z_{1,-1}+Z_{-1,1}+Z_{-1,-1},
\end{align}
and its full partition function
\be
\frac{1}{8}[\langle Z_{1,1}\rangle+\langle Z_{1,i}\rangle+\langle Z_{1,j}\rangle+\langle Z_{1,k}\rangle+\langle Z_{1,-1}\rangle].
\ee
These two notations for $Q_8$ relate as $(1,0)=i$, $(0,1)=j$ and $(1,1)=k$.  We can reduce back to the original group by declaring that $i$ acts trivially.  Then the remaining $\Z_2$ is generated by the equivalence class $g=\{j,-j,k,-k\}$.  With no modification to the action of $\Z_4$ in the $\Z_2$-twisted sectors, the $Q_8$ partition function becomes
\be
\frac{1}{8}[16\langle Z_{1,1}\rangle+8\langle Z_{1,g}\rangle]=\langle Z_{1,1}\rangle+2\cdot\frac{1}{2}[\langle Z_{1,1}\rangle+\langle Z_{1,g}\rangle],
\ee
yielding a single copy of the parent theory plus two copies of the $\Z_2$ orbifold.  If the original $\Z_2$ is anomalous, this result will be equally ill-defined.  If we take the quantum symmetry to be non-trivial, however, the $\langle Z_{1,j}\rangle$ and $\langle Z_{1,k}\rangle$ will cancel within themselves.  This leaves us with
\be
2\langle Z_{1,1}\rangle,
\ee
two copies of the parent theory.  While this non-central extension is not minimal, it is capable of curing the anomaly in the initial $\Z_2$ symmetry.  This will be our only non-central example, as in general we tend to find minimal extensions to be central.

\subsection{$G=\Z_2\times\Z_2$}
\label{z2z2}

In this section we tackle a more complicated example where the original symmetry is $\Z_2\times\Z_2$.  To start, we need to specify an anomaly in our symmetry, which entails a choice of $\omega\in H^3(\Z_2\times\Z_2,\U(1))\cong\Z_2^3$.  We'll write the group as $\Z_2\times\Z_2=\{1,a,b,c\}$ and for readability drop the arguments of $\om$ into subscripts, e.g.\ $\om(a,b,b)=\om_{abb}$.  Then a choice of representatives for each class in $H^3(\Z_2\times\Z_2,\U(1))$ is
\be
\begin{matrix}
	\label{z2z2omega}
	\om_{aaa}=\e_a, & \om_{aab}=1, & \om_{aac}=\e_a, & \om_{aba}=1, & \om_{abb}=1, \\ & \om_{abc}=1, & \om_{aca}=\e_a, & \om_{acb}=1, & \om_{acc}=\e_a, \\ \om_{baa}=1, & \om_{bab}=1, & \om_{bac}=1, & \om_{bba}=\e_a\e_b\e_c, & \om_{bbb}=\e_b, \\ & \om_{bbc}=\e_a\e_c, & \om_{bca}=\e_a\e_b\e_c, & \om_{bcb}=\e_b, & \om_{bcc}=\e_a\e_c, \\ \om_{caa}=\e_a, & \om_{cab}=1, & \om_{cac}=\e_a, & \om_{cba}=\e_a\e_b\e_c, & \om_{cbb}=\e_b, \\ & \om_{cbc}=\e_a\e_c, & \om_{cca}=\e_b\e_c, & \om_{ccb}=\e_b, & \om_{ccc}=\e_c,
\end{matrix}
\ee
where $\e_a^2=\e_b^2=\e_c^2=1$ label the cohomology class.  Then for $m=n_{aa}x_{aa}+\cdots+n_{cc}x_{cc}$, we have
\bea
(b(a))(m) &=& \e_a^{n_{aa}+n_{ac}+n_{ca}+n_{cc}},\\
(b(b))(m) &=& \lp\e_a\e_c\rp^{n_{ba}+n_{bc}+n_{ca}+n_{cc}}\e_b^{n_{ba}+n_{bb}+n_{ca}+n_{cb}},\\
(b(c))(m) &=& \e_a^{n_{aa}+n_{ac}+n_{ba}+n_{bc}}\e_b^{n_{ba}+n_{bb}+n_{ca}+n_{cb}}\e_c^{n_{ba}+n_{bc}+n_{ca}+n_{cc}}.
\eea

The orbit structure of a $\Z_2\times\Z_2$ orbifold is
\begin{align}
\langle Z_{1,1}\rangle&=Z_{1,1},\\
\langle Z_{1,a}\rangle&=Z_{1,a}+Z_{a,1}+Z_{a,a},\\
\langle Z_{1,b}\rangle&=Z_{1,b}+Z_{b,1}+Z_{b,b},\\
\langle Z_{1,c}\rangle&=Z_{1,c}+Z_{c,1}+Z_{c,c},\\
\langle Z_{a,b}\rangle&=Z_{a,b}+Z_{a,c}+Z_{b,a}+Z_{b,c}+Z_{c,a}+Z_{c,b}.
\end{align}
Its full partition function is, as given in (\ref{z2z2pf}), is
\be
\frac{1}{4}[\langle Z_{1,1}\rangle+\langle Z_{1,a}\rangle+\langle Z_{1,b}\rangle+\langle Z_{1,c}\rangle\pm \langle Z_{a,b}\rangle].
\ee
Employing (\ref{tanomalyspec}) and (\ref{z2z2omega}), we find the following anomalous transformations (there are others, but these are sufficient to describe the anomalies in each orbit):
\begin{align}
\label{z2z2anomaly}
Z_{a,1}(\tau+2)&=\epsilon_aZ_{a,1}(\tau)\\
Z_{b,1}(\tau+2)&=\epsilon_bZ_{b,1}(\tau)\\
Z_{c,1}(\tau+2)&=\epsilon_cZ_{c,1}(\tau)\\
Z_{a,b}(\tau+2)&=\epsilon_aZ_{a,b}(\tau)\\
Z_{b,c}(\tau+2)&=\epsilon_bZ_{b,c}(\tau)\\\label{z2z2anomalylast}
Z_{c,a}(\tau+2)&=\epsilon_cZ_{c,a}(\tau).
\end{align}
Thus the orbits $\langle Z_{1,a}\rangle$, $\langle Z_{1,b}\rangle$ and $\langle Z_{1,c}\rangle$ are anomalous when their respective $\epsilon$ are non-zero, and the disconnected orbit $\langle Z_{a,b}\rangle$ is anomalous for any non-trivial anomaly.

\subsubsection{$1\to\Z_2\to D_4\to\Z_2\times\Z_2\to 1$}

The first anomaly we will examine corresponds to $(\epsilon_a,\epsilon_b,\epsilon_c)=(1,1,-1)$.  In this case,
\be
\bar{A}=\{m|n_{ba}+n_{bc}+n_{ca}+n_{cc}\ \mathrm{even}\}.
\ee
Then $C=\{[0],[x_{cc}]\}\cong\Z_2$.  A short computation shows that the $G$ action on $C$ is trivial:
\be
a\cdot[x_{cc}]=[x_{bc}+x_{ac}]=[x_{cc}],\quad b\cdot[x_{cc}]=[x_{ac}+x_{bc}]=[x_{cc}],\quad c\cdot[x_{cc}]=[x_{cc}].
\ee
So we can compute the group multiplication in $\widehat{\G}$,
\be
\label{eq:ExampleGammaHatMultiplication}
([jx_{cc}],g)([kx_{cc}],h)=([(j+k)x_{cc}+x_{gh}],gh).
\ee
In particular we have $([0],c)^4=([0],a)^2=1$ and $([0],a)([0],c)([0],a)=([x_{cc}],c)=([0],c)^{-1}$, which lets us identify the group as $\widehat{\G}\cong D_4$, the dihedral group with eight elements.  The values of $b$ which could possibly be non-trivial are
\be
\label{d4b}
(b(a))([x_{cc}])=1 \quad (b(b))([x_{cc}])=-1 \quad (b(c))([x_{cc}])=-1.
\ee

We present the extended group $D_4$ as $\braket{r,s|r^4=s^2=(sr)^2=1}$.  This corresponds to the notation used above as $r=([0],c)$ and $s=([0],a)$.  The orbits of a $D_4$ orbifold are then
\begin{align}
\langle Z_{1,1}\rangle&=Z_{1,1},\\
\langle Z_{1,r}\rangle&=Z_{1,r}+Z_{1,r^3}+Z_{r,1}+Z_{r,r}+Z_{r,r^2}+Z_{r,r^3}+Z_{r_2,r}+Z_{r^2,r^3}\\\nonumber
&+Z_{r^3,1}+Z_{r^3,r}+Z_{r^3,r^2}+Z_{r^3,r^3},\\
\langle Z_{1,r^2}\rangle&=Z_{1,r^2}+Z_{r^2,1}+Z_{r^2,r^2},\\
\langle Z_{1,s}\rangle&=Z_{1,s}+Z_{s,1}+Z_{s,s},\\
\langle Z_{1,rs}\rangle&=Z_{1,rs}+Z_{rs,1}+Z_{rs,rs},\\
\langle Z_{1,r^2s}\rangle&=Z_{1,r^2s}+Z_{r^2s,1}+Z_{r^2s,r^2s},\\
\langle Z_{1,r^3s}\rangle&=Z_{1,r^3s}+Z_{r^3s,1}+Z_{r^3s,r^3s},\\
\langle Z_{r^2,s}\rangle&=Z_{r^2,s}+Z_{r^2,r^2s}+Z_{s,r^2}+Z_{s,r^2s}+Z_{r^2s,r^2}+Z_{r^2s,s},\\
\langle Z_{r^2,rs}\rangle&=Z_{r^2,rs}+Z_{r^2,r^3s}+Z_{rs,r^2}+Z_{rs,r^3s}+Z_{r^3s,r^2}+Z_{r^3s,rs},
\end{align}
and the full orbifold partition function is
\begin{multline}
\frac{1}{8}[\langle Z_{1,1}\rangle+\langle Z_{1,r}\rangle+\langle Z_{1,r^2}\rangle+\langle Z_{1,s}\rangle+\langle Z_{1,rs}\rangle+\langle Z_{1,r^2s}\rangle+\langle Z_{1,r^3s}\rangle\\
\pm\langle Z_{r^2,s}\rangle\pm\langle Z_{r^2,rs}\rangle].
\end{multline}

Using (\ref{bphases}) and (\ref{d4b}) to assign phases, we find that the $D_4$ orbits relate to those of $\Z_2\times\Z_2$ as
\begin{align}
\langle Z_{1,1}\rangle&\to \langle Z_{1,1}\rangle,\\
\langle Z_{1,r}\rangle&\to 0,\\
\langle Z_{1,r^2}\rangle&\to 3\langle Z_{1,1}\rangle,\\
\langle Z_{1,s}\rangle&\to \langle Z_{1,a}\rangle,\\
\langle Z_{1,rs}\rangle&\to \langle Z_{1,b}\rangle,\\
\langle Z_{1,r^2s}\rangle&\to \langle Z_{1,a}\rangle,\\
\langle Z_{1,r^3s}\rangle&\to \langle Z_{1,b}\rangle,\\
\langle Z_{r^2,s}\rangle&\to 2\langle Z_{1,a}\rangle,\\
\langle Z_{r^2,rs}\rangle&\to -2\langle Z_{1,b}\rangle.
\end{align}
This means that the full $D_4$ orbifold becomes
\be
\frac{1}{4}[2\langle Z_{1,1}\rangle+(1\pm 1)\langle Z_{1,a}\rangle+(1\mp 1)\langle Z_{1,b}\rangle].
\ee
Interestingly, we see that discrete torsion in the $D_4$ extension swaps us between two $\Z_2$ orbifolds -- one by $a$, the other by $b$.  Either way we have, as promised, a single copy of a non-anomalous orbifold, confirming that this was a minimal trivializing extension.

Why is it that, in this case, discrete torsion swaps between two trivializing extensions?  This example has $H^1(\Z_2\times\Z_2,\hat{\Z}_2)\cong\Z_2\times\Z_2$, which we will write as $\{1,x,y,z\}$.  Turning on discrete torsion in $D_4$ is equivalent to choosing a quantum symmetry belonging to one of the non-trivial classes in this $\Z_2\times\Z_2$ -- we'll say it's the class given by $x$.  Meanwhile, the quantum symmetry coming from $b$ as given by (\ref{d4b}) corresponds to a different non-trivial class -- call it the one given by $y$.  So turning on discrete torsion swaps us between group actions in the classes corresponding to $y$ and $z$.

This analysis further implies that there is a third trivializing extension which we have not considered: the same extension $1\to\Z_2\to D_4\to\Z_2\times\Z_2\to 1$, except this time the only quantum symmetry we turn on is the $D_4$ discrete torsion.  With the notation as above, a $D_4$ orbifold with discrete torsion decomposes to
\be
\frac{1}{2}[\langle Z_{1,1}\rangle+\langle Z_{1,c}\rangle],
\ee
a single copy of the $\Z_2$ orbifold by $c$.  This extension would trivialize any anomaly in $\Z_2\times\Z_2$ with $\epsilon_c=1$, and is our second example of curing an anomaly by turning on a quantum symmetry equivalent to discrete torsion (the first having appeared in section \ref{sec:z2anom}).

\subsubsection{$1\to\Z_2\times\Z_2\to\Z_4\rtimes\Z_4\to\Z_2\times\Z_2\to 1$}

A qualitatively different anomaly comes from the choice $(\epsilon_a,\epsilon_b,\epsilon_c)=(-1,-1,-1)$.  We see from (\ref{z2z2anomaly})-(\ref{z2z2anomalylast}) that, with this choice, every orbit besides $\langle Z_{1,1}\rangle$ is anomalous.  In this case we have
\be
\bar{A}=\{m|n_{aa}+n_{ac}+n_{ca}+n_{cc}\ \mathrm{and}\ n_{ba}+n_{bb}+n_{ca}+n_{cb}\ \mathrm{even}\},
\ee
\be
C=\{[0],[x_{aa}],[x_{bb}],[x_{aa}+x_{bb}]\}\cong\Z_2^2,
\ee
with trivial group action.  Under this idenficiation, $b$ becomes
\be
\begin{matrix}
(b(a))([x_{aa}])=-1, & (b(b))([x_{aa}])=1, & (b(c))([x_{aa}])=-1, \\
(b(a))([x_{bb}])=1, & (b(b))([x_{bb}])=-1, & (b(c))([x_{bb}])=-1, \\
(b(a))([x_{cc}])=-1, & (b(b))([x_{cc}])=-1, & (b(c))([x_{cc}])=1.
\end{matrix}
\ee

A computation reveals that $\widehat{\G}$ is the nontrivial semidirect product $\Z_4\ltimes\Z_4$.  This group has the presentation $\braket{x,y|x^4=y^4=1, [x,y]=x^2}$.  We'll choose to write the sixteen group elements as $x^iy^j$, $(i,j)\in \{0,1,2,3\}^2$.  We can relate this to the above notation as
\be
x=a, \quad y=b, \quad xy=c, \quad x^2=[x_{aa}], \quad y^2=[x_{bb}], \quad x^2y^2=[x_{aa}+x_{bb}].
\ee
The modular orbits of $\Z_4\rtimes\Z_4$ are
\begin{align}
	\langle Z_{1,1}\rangle&=Z_{1,1},\\
	\langle Z_{1,x}\rangle&=Z_{1,x}+Z_{1,x^3}+Z_{x,1}+Z_{x,x}+Z_{x,x^2}+Z_{x,x^3}+Z_{x^2,x}+Z_{x^2,x^3}\\\nonumber
	&+Z_{x^3,1}+Z_{x^3,x}+Z_{x^3,x^2}+Z_{x^3,x^3},\\
	\langle Z_{1,x^2}\rangle&=Z_{1,x^2}+Z_{x^2,1}+Z_{x^2,x^2},\\
	\langle Z_{1,y}\rangle&=Z_{1,y}+Z_{1,y^3}+Z_{y,1}+Z_{y,y}+Z_{y,y^2}+Z_{y,y^3}+Z_{y^2,y}+Z_{y^2,y^3}\\\nonumber
	&+Z_{y^3,1}+Z_{y^3,y}+Z_{y^3,y^2}+Z_{y^3,y^3},\\
	\langle Z_{1,y^2}\rangle&=Z_{1,y^2}+Z_{y^2,1}+Z_{y^2,y^2},\\
	\langle Z_{1,xy}\rangle&=Z_{1,xy}+Z_{1,xy^3}+Z_{y^2,xy}+Z_{y^2,xy^3}+Z_{xy,1}+Z_{xy,y^2}+Z_{xy,xy}\\\nonumber
	&+Z_{xy,xy^3}+Z_{xy^3,1}+Z_{xy^3,y^2}+Z_{xy^3,xy}+Z_{xy^3,xy^3},\\
	\langle Z_{1,x^2y}\rangle&=Z_{1,x^2y}+Z_{1,x^2y^3}+Z_{y^2,x^2y}+Z_{y^2,x^2y^3}+Z_{x^2y,1}+Z_{x^2y,y^2}+Z_{x^2y,x^2y}\\\nonumber
	&+Z_{x^2y,x^2y^3}+Z_{x^2y^3,1}+Z_{x^2y^3,y^2}+Z_{x^2y^3,x^2y}+Z_{x^2y^3,x^2y^3},\\
	\langle Z_{1,x^3y}\rangle&=Z_{1,x^3y}+Z_{1,x^3y^3}+Z_{y^2,x^3y}+Z_{y^2,x^3y^3}+Z_{x^3y,1}+Z_{x^3y,y^2}+Z_{x^3y,x^3y}\\\nonumber
	&+Z_{x^3y,x^3y^3}+Z_{x^3y^3,1}+Z_{x^3y^3,y^2}+Z_{x^3y^3,x^3y}+Z_{x^3y^3,x^3y^3},\\
	\langle Z_{1,xy^2}\rangle&=Z_{1,xy^2}+Z_{1,x^3y^2}+Z_{x^2,xy^2}+Z_{x^2,x^3y^2}+Z_{xy^2,1}+Z_{xy^2,x^2}+Z_{xy^2, xy^2}\\\nonumber
	&+Z_{xy^2,x^3y^2}+Z_{x^3y^2,1}+Z_{x^3y^2,x^2}+Z_{x^3y^2,xy^2}+Z_{x^3y^2,x^3y^2},\\
	\langle Z_{1,x^2y^2}\rangle&=Z_{1,x^2y^2}+Z_{x^2y^2,1}+Z_{x^2y^2,x^2y^2},\\
	\langle Z_{x,y^2}\rangle&=Z_{x,y^2}+Z_{x,xy^2}+Z_{x,x^2y^2}+Z_{x,x^3y^2}+Z_{y^2,x^3}+Z_{y^2,x^3y^2}+Z_{y^2,xy^2}\\\nonumber
	&+Z_{y^2,x}+Z_{xy^2,x^3}+Z_{xy^2,x^2y^2}+Z_{xy^2,x}+Z_{xy^2,y^2}+Z_{x^2y^2,x}+Z_{x^2y^2,x^3}\\\nonumber
	&+Z_{x^2y^2,xy^2}+Z_{x^2y^2,x^3y^2}+Z_{x^3y^2,x^3}+Z_{x^3y^2,y^2}+Z_{x^3y^2,x}+Z_{x^3y^2,x^2y^2}\\\nonumber
	&+Z_{x^3,xy^2}+Z_{x^3,x^2y^2}+Z_{x^3,x^3y^2}+Z_{x^3,y^2},\\
	\langle Z_{y,x^2}\rangle&=Z_{y,x^2}+Z_{y,x^2y}+Z_{y,x^2y^2}+Z_{y,x^2y^3}+Z_{x^2,y^3}+Z_{x^2,x^2y^3}+Z_{x^2,x^2y}\\\nonumber
	&+Z_{x^2,y}+Z_{x^2y^3,y^3}+Z_{x^2y^3,x^2}+Z_{x^2y^3,y}+Z_{x^2y^3,x^2y^2}+Z_{x^2y^2,y}+Z_{x^2y^2,y^3}\\\nonumber
	&+Z_{x^2y^2,x^2y}+Z_{x^2y^2,x^2y^3}+Z_{x^2y^3,y^3}+Z_{x^2y^3,x^2}+Z_{x^2y^3,y}+Z_{x^2y^3,x^2y^2}\\\nonumber
	&+Z_{y^3,x^2}+Z_{y^3,x^2y}+Z_{y^3,x^2y^2}+Z_{y^3,x^2y^3},\\
	\langle Z_{xy,x^2y^2}\rangle&=Z_{xy,x^2y^2}+Z_{xy,x^3y}+Z_{xy,x^2}+Z_{xy,x^3y^3}+Z_{x^2y^2,xy}+Z_{x^2y^2,x^3y^3}\\\nonumber
	&+Z_{x^2y^2,xy^3}+Z_{x^2y^2,x^3y}+Z_{x^2,xy}+Z_{x^2,x^3y}+Z_{x^2,xy^3}+Z_{x^2,x^3y^3}\\\nonumber
	&+Z_{xy^3,x^2y^2}+Z_{xy^3,x^3y^3}+Z_{xy^3,x^2}+Z_{xy^3,x^3y}+Z_{x^3y,xy^3}+Z_{x^3y,x^2y^2}\\\nonumber
	&+Z_{x^3y,xy}+Z_{x^3y,x^2}+Z_{x^3y^3,x^2}+Z_{x^3y^3,xy}+Z_{x^3y^3,x^2y^2}+Z_{x^3y^3,xy^3},\\
	\langle Z_{x^2,y^2}\rangle&=Z_{x^2,y^2}+Z_{x^2,x^2y^2}+Z_{y^2,x^2}+Z_{y^2,x^2y^2}+Z_{x^2y^2,x^2}+Z_{x^2y^2,y^3}.
\end{align}
The full orbifold partition function is then
\begin{multline}
\frac{1}{16}[Z_{1,1}+\langle Z_{1,x}\rangle+\langle Z_{1,x^2}\rangle+\langle Z_{1,y}\rangle+\langle Z_{1,y^2}\rangle+\langle Z_{1,xy}\rangle+\langle Z_{1,x^2y}\rangle+\langle Z_{1,x^3y}\rangle+\langle Z_{1,xy^2}\rangle\\
+\langle Z_{1,x^2y^2}\rangle+\langle Z_{x,y^2}\rangle\pm\langle Z_{y,x^2}\rangle\pm\langle Z_{xy,x^2y^2}\rangle+\langle Z_{x^2,y^2}\rangle].
\end{multline}
Calculating coefficients of decomposition from $b$, we find that every orbit that does not reduce to the parent theory partition function vanishes.  Sixteen of the $\Z_4\rtimes\Z_4$ partial traces reduce to $Z_{1,1}$, so we end up with a single copy of the parent theory, as we should have expected from a minimal trivializing extension when all of the other orbits were anomalous.

\subsection{$G=\Z_2\times\Z_4$}

In this section we examine anomalies in the group $\Z_2\times\Z_4$, which will be generated by $a^2=b^4=1$.  The orbits of a $\Z_2\times\Z_4$ orbifold are then
\begin{align}
	\langle Z_{1,1}\rangle&=Z_{1,1},\\
	\langle Z_{1,a}\rangle&=Z_{1,a}+Z_{a,1}+Z_{a,a},\\
	\langle Z_{1,b}\rangle&=Z_{1,b}+Z_{1,b^3}+Z_{b,1}+Z_{b,b^3}+Z_{b,b^2}+Z_{b,b}+Z_{b^3,b^3}+Z_{b^3,1}+Z_{b^3,b}\\\nonumber
	&+Z_{b^3,b^2}+Z_{b^2,b^3}+Z_{b^2,b},\\
	\langle Z_{1,b^2}\rangle&=Z_{1,b^2}+Z_{b^2,1}+Z_{b^2,b^2},\\
	\langle Z_{1,ab}\rangle&=Z_{1,ab}+Z_{1,ab^3}+Z_{ab,1}+Z_{ab,ab}+Z_{ab,b^2}+Z_{ab,ab^3}+Z_{b^2,ab^3}+Z_{b^2,ab}\\\nonumber&+Z_{ab^3,ab^3}+Z_{ab^3,1}+Z_{ab^3,ab}+Z_{ab^3,b^2},\\
	\langle Z_{1,ab^2}\rangle&=Z_{1,ab^2}+Z_{ab^2,1}+Z_{ab^2,ab^2},\\
	\langle Z_{a,b}\rangle&=Z_{a,b}+Z_{a,b^3}+Z_{a,ab}+Z_{a,ab^3}+Z_{b,a}+Z_{b,ab}+Z_{b,ab^2}+Z_{b,ab^3}+Z_{b^3,ab^3},\\\nonumber&+Z_{b^3,a}+Z_{b^3,ab}+Z_{b^3,ab^2}+Z_{ab,b^3}+Z_{ab,ab^2}+Z_{ab,b}+Z_{ab,a}+Z_{ab^2,b}\\\nonumber&+Z_{ab^2,b^3}+Z_{ab^2,ab}+Z_{ab^2,ab^3}+Z_{ab^3,b^3}+Z_{ab^3,a}+Z_{ab^3,b}+Z_{ab^3,ab^2},\\
	\langle Z_{a,b^2}\rangle&=Z_{a,b^2}+Z_{a,ab^2}+Z_{b^2,a}+Z_{b^2,ab^2}+Z_{ab^2,a}+Z_{ab^2,b^2},
\end{align}
while the full orbifold partition function is given by
\be
\frac{1}{8}[Z_{1,1}+\langle Z_{1,a}\rangle+\langle Z_{1,b}\rangle+\langle Z_{1,b^2}\rangle+\langle Z_{1,ab}\rangle+\langle Z_{1,ab^2}\rangle\pm\langle Z_{a,b}\rangle+\langle Z_{a,b^2}\rangle].
\ee
Making a particular choice of representatives of $H^3(\Z_2\times\Z_4,U(1))$, the anomalous phases that appear in the untwisted orbits are given by
\begin{align}
Z_{a,1}(\tau+2)&=\gamma Z_{a,1}(\tau)\\
Z_{b,1}(\tau+4)&=\alpha^3 Z_{b,1}(\tau)\\
Z_{b^2,1}(\tau+2)&=\alpha^2Z_{b^2,1}(\tau)\\
Z_{ab,1}(\tau+4)&=\alpha^3\beta Z_{ab,1}(\tau)\\
Z_{ab^2,1}(\tau+2)&=\alpha^2\beta\gamma Z_{ab^2,1}(\tau)
\end{align}
where $\alpha^4=\beta^2=\gamma^2=1$ determine the cohomology class in $H^3(\Z_2\times\Z_4,U(1))\cong\Z_4\times\Z_2\times\Z_2$.

\subsubsection{$1\to\Z_2\to\Z_2\times\Z_8\to\Z_2\times\Z_4\to 1$}

To begin with, we examine the case $\alpha=-1$, $\beta=\gamma=1$.  Intuitively this is the case where $b$ is anomalous, and we have a non-anomalous $\Z_2\times\Z_2$ subgroup generated by $a$ and $b^2$.  This is essentially equivalent to a $\Z_4$ with an order two anomaly, which we would cure by extending to $\Z_8$.  In the present situation the $\Z_2$ generated by $a$ is a bystander and we can extend to $\Z_2\times\Z_8$ to cure the anomaly.  Choosing to write the extending $\Z_2$ as $b^4$, we have the $\Z_2\times\Z_8$ orbifold partition function
\begin{multline}
\label{z2z8}
\frac{1}{16}[\langle Z_{1,1}\rangle+\langle Z_{1,b}\rangle+\langle Z_{1,b^2}\rangle+\langle Z_{1,b^4}\rangle+\langle Z_{1,a}\rangle+\langle Z_{1,ab}\rangle\\
+\langle Z_{1,ab^2}\rangle+\langle Z_{1,ab^4}\rangle\pm\langle Z_{a,b}\rangle+\langle Z_{a,b^2}\rangle+\langle Z_{a,b^4}\rangle].
\end{multline}
In projecting back to $\Z_2\times\Z_4$ the coefficients of decomposition are
\begin{align}
\langle Z_{1,1}\rangle&\to\langle Z_{1,1}\rangle,\\
\langle Z_{1,b}\rangle&\to 0,\\
\langle Z_{1,b^2}\rangle&\to 4\langle Z_{1,b^2}\rangle,\\
\langle Z_{1,b^4}\rangle&\to 3\langle Z_{1,1}\rangle,\\
\langle Z_{1,a}\rangle&\to \langle Z_{1,a}\rangle,\\
\langle Z_{1,ab}\rangle&\to 0,\\
\langle Z_{1,ab^2}\rangle&\to 4\langle Z_{1,ab^2}\rangle,\\
\langle Z_{1,ab^4}\rangle&\to\langle Z_{1,a}\rangle,\\
\langle Z_{a,b}\rangle&\to 0,\\
\langle Z_{a,b^2}\rangle&\to 4\langle Z_{a,b^2}\rangle,\\
\langle Z_{a,b^4}\rangle&\to 2\langle Z_{1,a}\rangle.
\end{align}
With these relations, the $\Z_2\times\Z_8$ partition function (\ref{z2z8}) becomes
\be
\frac{1}{4}[\langle Z_{1,1}\rangle+\langle Z_{1,a}\rangle+\langle Z_{1,b^2}\rangle+\langle Z_{1,ab^2}\rangle+\langle Z_{a,b^2}\rangle],
\ee
which is the orbifold partition function for the non-anomalous $\Z_2\times\Z_2$ subgroup generated by $a$ and $b^2$, with the trivial choice of discrete torsion.

\subsubsection{$1\to\Z_4\to\Z_4\times\Z_8\to\Z_2\times\Z_4\to 1$}
\label{Z4xZ8}

If we were to choose $\beta=-1$, $\alpha=\gamma=1$, the non-anomalous subgroups would be the $\Z_2$ generated by $a$ and the $\Z_4$ generated by $b$ (but not any mix of the two).  This is the first situation in which we have distinct non-anomalous subgroups of differing sizes.  In order to build a trivializing extension, we choose our $H^3(\Z_2\times\Z_4,U(1))$ representative to be
\be
\om(a^sb^x,a^pb^y,a^qb^z)=i^{x(pq+2yz)}.
\ee
One can check from (\ref{tanomalyspec}) that this cocycle indeed produces $\beta=-1,\alpha=\gamma=1$.  When we reduce Tachikawa's construction in this example, we find
\be
C=\{[0],[x_{a,a}],[x_{b,b}],[x_{ab,ab}]\}\cong\Z_4.
\ee
The resulting $\widehat{\Gamma}$ turns out to be $\Z_4\times\Z_8$, which we can take to be generated by $([0],a)^8=([0],b)^4=1$.  Omitting details of the calculation (as there are now over one thousand partial traces involved), the $\Z_4\times\Z_8$ orbifold reduces to
\be
\frac{1}{2}[\langle Z_{1,1}\rangle+\langle Z_{1,a}\rangle],
\ee
which is the orbifold by the non-anomalous $\Z_2$ subgroup generated by $a$.  We expect that there should exist a different choice of representative for $\omega$ which produces a $\Z_2$ extension reducing to the $\Z_4$ orbifold generated by $b$, though we have not identified it.

\section{Discussion and Conclusions}
\label{sec:conclusion}

Through numerous examples we have seen that an anomalous symmetry $G$ can be extended to a larger group $\Gamma$ such that the orbifold by $\Gamma$ is equivalent to a direct sum of orbifolds by non-anomalous subgroups of $G$ and is therefore consistent.  In each example we examined, the ability to find such a resolution depended on the existence of non-trivial quantum symmetries given by elements of $H^1(G,\hat{K})$ which encode the action of the extending group $K$ on the $G$-twisted states.  Sometimes this modification is equivalent to turning on discrete torsion, while other times it is a choice that is unique to orbifolds with trivially-acting subgroups.

While we have shown by construction that there exists at least one such extension for each symmetry $G$ and anomaly $\om\in H^3(G,U(1))$, we could aspire to even more specific control over the form of the extension.  Specifically, we could ask the following: given a group $G$ and a subgroup $H$ of $G$, does there exist an extension $\Gamma$ of $G$ such that the orbifold by $\Gamma$ is equivalent to a single copy of the orbifold by $H$?  This would imply that we can find a minimal trivializing extension that produces an orbifold equivalent to the orbifold by any non-anomalous subgroup of $G$.

In addition, we could hope for a more precise characterization of the decomposition at play.  In particular this should include a general form for the $\Gamma$ orbifold's decomposition in the spirit of \cite{decomp4}, the precise relation between discrete torsion and quantum symmetries, and a description of the open string sector related to such extensions.  We are currently investigating such questions and hope to begin to tackle them in \cite{decomp_qs} and \cite{decompanomalies}.

\section*{Acknowledgements}

We would like to thank T.~Pantev for useful discussions.  
D.R. was partially supported by 
NSF grant PHY-1820867.
E.S. was partially supported by NSF grants
PHY-1720321
and PHY-2014086.

\appendix
\section{Non-Effective Extensions to Non-Abelian Groups}
\label{appendix:nonabelian}

In this appendix we work through the example of extending a $\Z_2\times\Z_2$ symmetry by a non-effectively acting $\Z_2$ to the group of unit quaternions, $Q_8$, which is non-abelian.  We will see how and why the degeneration process of non-abelian orbifolds with trivially-acting subgroups is modified, then examine the inclusion of quantum symmetries in such orbifolds.  We restate a number of formulae from the main text (mainly the orbit structures of $\Z_2\times\Z_2$ and $Q_8$ orbifolds) to make the appendix more self-contained.

\subsection{The $Q_8$ Orbifold}

We present the group $Q_8$ as $\braket{i,j,k|i^2=j^2=k^2=ijk=-1}$.  Its genus one modular orbits are
\begin{align}
\langle Z_{1,1}\rangle&=Z_{1,1},\\
\langle Z_{1,-1}\rangle&=Z_{1,-1}+Z_{-1,1}+Z_{-1,-1},\\
\langle Z_{1,i}\rangle&=Z_{1,i}+Z_{1,-i}+Z_{i,1}+Z_{-i,1}+Z_{i,i}+Z_{i,-1}\\\notag
&+Z_{i,-i}+Z_{-i,i}+Z_{-i,-1}+Z_{-i,-i}+Z_{-1,i}+Z_{-1,-i},\\
\langle Z_{1,j}\rangle&=Z_{1,j}+Z_{1,-j}+Z_{j,1}+Z_{-j,1}+Z_{j,j}+Z_{j,-1}\\\notag
&+Z_{j,-j}+Z_{-j,j}+Z_{-j,-1}+Z_{-j,-j}+Z_{-1,j}+Z_{-1,-j},\\
\langle Z_{1,k}\rangle&=Z_{1,k}+Z_{1,-k}+Z_{k,1}+Z_{-k,1}+Z_{k,k}+Z_{k,-1}\\\notag
&+Z_{k,-k}+Z_{-k,k}+Z_{-k,-1}+Z_{-k,-k}+Z_{-1,k}+Z_{-1,-k},
\end{align}
and its full torus partition function is
\be
\label{Q8pf}
\frac{1}{8}[\langle Z_{1,1}\rangle+\langle Z_{1,-1}\rangle+\langle Z_{1,i}\rangle+\langle Z_{1,j}\rangle+\langle Z_{1,k}\rangle].
\ee
Note that there is no choice of discrete torsion in this orbifold -- we could see this without even calculating $H^2(Q_8,U(1))$ (which is in fact trivial) by noting that all of the orbits include an untwisted sector partial trace, so all of the relative coefficients are fixed by modular transformations.  Now we examine the Sp$(4;\Z)$ orbits of the genus two partial traces $Z_{g_1,g_2,h_1,h_2}$ with $g_1h_1g_1^{-1}h_1^{-1}g_2h_2g_2^{-1}h_2^{-1}=[g_1,h_1][g_2,h_2]=1$, which are
\begin{itemize}
	\item $\langle Z_{1,1,1,1}\rangle$ with 1 partial trace,
	\item $\langle Z_{1,1,1,-1}\rangle$ with 15 partial traces,
	\item $\langle Z_{1,1,1,i}\rangle$ with 240 partial traces,
	\item $\langle Z_{1,1,1,j}\rangle$ with 240 partial traces,
	\item $\langle Z_{1,1,1,k}\rangle$ with 240 partial traces and
	\item $\langle Z_{1,1,i,j}\rangle$ with 1440 partial traces.
\end{itemize}
They form the full genus two partition function as
\be
\label{Q8pfg2}
\frac{1}{8^2}[\langle Z_{1,1,1,1}\rangle+\langle Z_{1,1,1,-1}\rangle+\langle Z_{1,1,1,i}\rangle+\langle Z_{1,1,1,j}\rangle+\langle Z_{1,1,1,k}\rangle+\langle Z_{1,1,i,j}\rangle].
\ee
Note that the orbit  $\langle Z_{1,1,i,j}\rangle$ includes partial traces such as $Z_{i,i,j,j}$ which has $[g_1,h_1]=[g_2,h_2]=-1$.  Under a degeneration that takes the genus two surface to a pair of genus one surfaces connected by a thin tube, these partial traces have a TDL labeled by the element $-1$ running through the tube.  We can insert a sum over a complete set of $-1$-twisted states in the tube, replacing the genus two partial trace by a weighted sum of genus one one-point functions.  If $-1$ acts effectively on the parent theory, then all of these states have positive conformal weights and in the degeneration limit they will be subleading to terms where $[g_1,h_1]=[g_2,h_2]=1$ and we can have the untwisted vacuum state propagating through the tube.

In total, the genus two $Q_8$ partial traces degenerate to genus one as (the convention here, as in section \ref{sec:actions}, is that the ordering of the genus one orbits determines the torus on which they reside)
\begin{align}
\langle Z_{1,1,1,1}\rangle\to& \langle Z_{1,1}\rangle\langle Z_{1,1}\rangle,\\
\langle Z_{1,1,1,-1}\rangle\to& \langle Z_{1,1}\rangle\langle Z_{1,-1}\rangle+\langle Z_{1,-1}\rangle\langle Z_{1,1}\rangle+\langle Z_{1,-1}\rangle\langle Z_{1,-1}\rangle,\\
\langle Z_{1,1,1,i}\rangle\to& \langle Z_{1,1}\rangle\langle Z_{1,i}\rangle+\langle Z_{1,i}\rangle\langle Z_{1,1}\rangle+\langle Z_{1,-1}\rangle\langle Z_{1,i}\rangle\\\notag
+&\langle Z_{1,i}\rangle\langle Z_{1,-1}\rangle+\langle Z_{1,i}\rangle\langle Z_{1,i}\rangle,\\
\langle Z_{1,1,1,j}\rangle\to& \langle Z_{1,1}\rangle\langle Z_{1,j}\rangle+\langle Z_{1,j}\rangle\langle Z_{1,1}\rangle+\langle Z_{1,-1}\rangle\langle Z_{1,j}\rangle\\\notag
+&\langle Z_{1,j}\rangle\langle Z_{1,-1}\rangle+\langle Z_{1,j}\rangle\langle Z_{1,j}\rangle,\\
\langle Z_{1,1,1,k}\rangle\to& \langle Z_{1,1}\rangle\langle Z_{1,k}\rangle+\langle Z_{1,k}\rangle\langle Z_{1,1}\rangle+\langle Z_{1,-1}\rangle\langle Z_{1,k}\rangle\\\notag
+&\langle Z_{1,k}\rangle\langle Z_{1,-1}\rangle+\langle Z_{1,k}\rangle\langle Z_{1,k}\rangle,\\
\langle Z_{1,1,i,j}\rangle\to& \langle Z_{1,i}\rangle\langle Z_{1,j}\rangle+\langle Z_{1,j}\rangle\langle Z_{1,i}\rangle+\langle Z_{1,i}\rangle\langle Z_{1,k}\rangle\\\notag
+&\langle Z_{1,k}\rangle\langle Z_{1,i}\rangle+\langle Z_{1,j}\rangle\langle Z_{1,k}\rangle+\langle Z_{1,k}\rangle\langle Z_{1,j}\rangle\\\notag
+&\langle Z_{i,j}[-1]\rangle\langle Z_{i,j}[-1]\rangle,
\end{align}
where in the last line, $\langle Z_{i,j}[-1]\rangle$ represents a genus one amplitude with TDLs labeled by $i$ and $j$ wrapping the two cycles of the torus and a TDL labeled by $-1$ leaving their crossing point and ending on the insertion of the lowest weight state in the $-1$-twisted sector.  If $-1$ acts effectively, then this operator has positive conformal weight and this term will not contribute to leading order in the degeneration.  However, if $-1$ does not act effectively then there will be a weight-zero state in the $-1$-twisted sector and this last term will contribute.  This should be a general feature of orbifolds with non-effective subgroups -- trivially-acting, non-identity elements in the commutator subgroup will lead to these additional contributions.

In the effective case, where we don't need to keep that last term, degeneration yields a copy of the genus one partition function (\ref{Q8pf}) on each torus, as we would expect.

\subsection{Decomposition}
\label{app:decomp}

$Q_8$ can be cast as a central extension $1\to K=\Z_2\to \Gamma =Q_8\to G =\Z_2\times\Z_2\to 1$, where the extending group $K$ is generated by the element $-1$.  We write $G=\Z_2\times\Z_2=\{1,a,b,c\}$ such that modding out by $-1$ sends $\{1,-1\}$ to $1$, $\{i,-i\}$ to $a$, $\{j,-j\}$ to $b$ and $\{k,-k\}$ to $c$.  As a reminder ,the genus one $\Z_2\times\Z_2$ orbits are
\begin{align}
\langle Z_{1,1}\rangle&=Z_{1,1},\\
\langle Z_{1,a}\rangle&=Z_{1,a}+Z_{a,1}+Z_{a,a},\\
\langle Z_{1,b}\rangle&=Z_{1,b}+Z_{b,1}+Z_{b,b},\\
\langle Z_{1,c}\rangle&=Z_{1,c}+Z_{c,1}+Z_{c,c},\\
\langle Z_{a,b}\rangle&=Z_{a,b}+Z_{a,c}+Z_{b,a}+Z_{b,c}+Z_{c,a}+Z_{c,b},
\end{align}
with the full partition function being
\be
Z_\pm=\frac{1}{4}[\langle Z_{1,1}\rangle+\langle Z_{1,a}\rangle+\langle Z_{1,b}\rangle+\langle Z_{1,c}\rangle\pm \langle Z_{a,b}\rangle].
\ee
We can see the $\Z_2$ discrete torsion arising as the coefficient of the disconnected orbit $\langle Z_{a,b}\rangle$.  At genus two there are six $\Z_2\times\Z_2$ orbits:
\begin{itemize}
	\item $\langle Z_{1,1,1,1}\rangle$ with 1 partial trace,
	\item $\langle Z_{1,1,1,a}\rangle$ with 15 partial traces,
	\item $\langle Z_{1,1,1,b}\rangle$ with 15 partial traces,
	\item $\langle Z_{1,1,1,c}\rangle$ with 15 partial traces,
	\item $\langle Z_{1,1,a,b}\rangle$ with 90 partial traces and
	\item $\langle Z_{1,a,1,b}\rangle$ with 120 partial traces.
\end{itemize}
These combine into the full partition function as
\be
\frac{1}{4^2}[\langle Z_{1,1,1,1}\rangle+\langle Z_{1,1,1,a}\rangle+\langle Z_{1,1,1,b}\rangle+\langle Z_{1,1,1,c}\rangle+\langle Z_{1,1,a,b}\rangle\pm \langle Z_{1,a,1,b}\rangle].
\ee

The genus two $\Z_2\times\Z_2$ partial traces degenerate to genus one as
\begin{align}
\langle Z_{1,1,1,1}\rangle\to&\langle Z_{1,1}\rangle\langle Z_{1,1}\rangle,\\
\langle Z_{1,1,1,a}\rangle\to&\langle Z_{1,1}\rangle\langle Z_{1,a}\rangle+\langle Z_{1,a}\rangle\langle Z_{1,1}\rangle+\langle Z_{1,a}\rangle\langle Z_{1,a}\rangle,\\
\langle Z_{1,1,1,b}\rangle\to&\langle Z_{1,1}\rangle\langle Z_{1,b}\rangle+\langle Z_{1,b}\rangle\langle Z_{1,1}\rangle+\langle Z_{1,b}\rangle\langle Z_{1,b}\rangle,\\
\langle Z_{1,1,1,a}\rangle\to&\langle Z_{1,1}\rangle\langle Z_{1,c}\rangle+\langle Z_{1,c}\rangle\langle Z_{1,1}\rangle+\langle Z_{1,c}\rangle\langle Z_{1,c}\rangle,\\
\langle Z_{1,1,a,b}\rangle\to&\langle Z_{1,a}\rangle\langle Z_{1,b}\rangle+\langle Z_{1,b}\rangle\langle Z_{1,a}\rangle+\langle Z_{1,b}\rangle\langle Z_{1,c}\rangle\\\notag
+&\langle Z_{1,c}\rangle\langle Z_{1,b}\rangle+\langle Z_{1,c}\rangle\langle Z_{1,a}\rangle+\langle Z_{1,a}\rangle\langle Z_{1,c}\rangle\\\notag
+&\langle Z_{a,b}\rangle\langle Z_{a,b}\rangle,\\
\langle Z_{1,a,1,b}\rangle\to&\langle Z_{1,1}\rangle\langle Z_{a,b}\rangle+\langle Z_{a,b}\rangle\langle Z_{1,1}\rangle+\langle Z_{1,a}\rangle\langle Z_{a,b}\rangle\\\notag
+&\langle Z_{a,b}\rangle\langle Z_{1,a}\rangle+\langle Z_{1,b}\rangle\langle Z_{a,b}\rangle+\langle Z_{a,b}\rangle\langle Z_{1,b}\rangle\\\notag
+&\langle Z_{1,c}\rangle\langle Z_{a,b}\rangle+\langle Z_{a,b}\rangle\langle Z_{1,c}\rangle.
\end{align}

Now if $K$ acts non-effectively, then the genus one $Q_8$ partial traces decompose to $\Z_2\times\Z_2$ partial traces as
\begin{align}
\langle Z_{1,1}\rangle&\to\langle Z_{1,1}\rangle,\\
\langle Z_{1,-1}\rangle&\to 3\langle Z_{1,1}\rangle,\\
\langle Z_{1,i}\rangle&\to 4\langle Z_{1,a}\rangle,\\
\langle Z_{1,j}\rangle&\to 4\langle Z_{1,b}\rangle,\\
\langle Z_{1,k}\rangle&\to 4\langle Z_{1,c}\rangle,
\end{align}
and we also have
\be
\langle Z_{i,j}[-1]\rangle\to 4\langle Z_{a,b}\rangle.
\ee
At genus two we have
\begin{align}
\langle Z_{1,1,1,1}\rangle&\to\langle Z_{1,1,1,1}\rangle,\\
\langle Z_{1,1,1,-1}\rangle&\to 15\langle Z_{1,1,1,1}\rangle,\\
\langle Z_{1,1,1,i}\rangle&\to 16\langle Z_{1,1,1,a}\rangle,\\
\langle Z_{1,1,1,j}\rangle&\to 16\langle Z_{1,1,1,b}\rangle,\\
\langle Z_{1,1,1,k}\rangle&\to 16\langle Z_{1,1,1,c}\rangle,\\
\langle Z_{1,1,i,j}\rangle&\to 16\langle Z_{1,1,a,b}\rangle.
\end{align}
Applying these decompositions to the full $Q_8$ partition function at genus one gives
\be
\label{ztilde}
\frac{1}{2}[\langle Z_{1,1}\rangle+\langle Z_{1,a}\rangle+\langle Z_{1,b}\rangle+\langle Z_{1,c}\rangle]=Z_++Z_-,
\ee
which is the direct sum of two $\Z_2\times\Z_2$ partition functions with opposite choices of discrete torsion, such that the disconnected orbits have canceled out.  At genus two decomposition gives us
\be
\label{q8decompg2}
\frac{1}{2^2}[\langle Z_{1,1,1,1}\rangle+\langle Z_{1,1,1,a}\rangle+\langle Z_{1,1,1,b}\rangle+\langle Z_{1,1,1c}\rangle+\langle Z_{1,1,a,b}\rangle].
\ee

Now whether we apply decomposition first and then degeneration, or degeneration first and then decomposition, the genus two $Q_8$ partition function becomes
\be
(Z_++Z_-)(Z_++Z_-)+(Z_+-Z_-)(Z_+-Z_-)=2Z_+Z_++2Z_-Z_-.
\ee
On the lhs we see the first term is the `usual' degeneration result (the genus one partition function on each torus) while the second term is the contribution from the non-effective weight-zero state $-1$.  The fact that this expression simplifies to something diagonal matches the expectation for a disjoint union theory, while the factor of two comes from the existence of two weight-zero states in the orbifold theory.  Note that the terms involving $\langle Z_{i,j}[-1]\rangle$ were necessary for the two possible orderings of degeneration and decomposition to agree.

%

\subsection{Quantum Symmetries}

Let us turn on a quantum symmetry in the $Q_8$ orbifold.  The possible quantum symmetries are classified by $H^1(\Z_2\times\Z_2,\hat{\Z}_2)\cong\Z_2\times\Z_2$.  The corresponding phases enter the partition function under degeneration as
\be
Z_{(k_1,g_1),(k_2,g_2)}\to \frac{B(k_2,g_1)}{B(k_1,g_2)}Z_{g_1,g_2}
\ee
where $B(k,g)$ is a homomorphism in both of its arguments and is normalized.  Therefore to specify it all we need to know are the values $\alpha=B(1,a)$, $\beta=B(1,b)$ and $\alpha\beta=B(1,c)$, with $\alpha^2=\beta^2=1$. Using this notation, the genus one decomposition relations become
\begin{align}
\langle Z_{1,1}\rangle&\to\langle Z_{1,1}\rangle,\\
\langle Z_{1,-1}\rangle&\to 3\langle Z_{1,1}\rangle,\\
\langle Z_{1,i}\rangle&\to 2(1+\alpha)\langle Z_{1,a}\rangle,\\
\langle Z_{1,j}\rangle&\to 2(1+\beta)\langle Z_{1,b}\rangle,\\
\langle Z_{1,k}\rangle&\to 2(1+\alpha\beta)\langle Z_{1,c}\rangle,\\
\langle Z_{i,j}[-1]\rangle&\to (1+\alpha+\beta+\alpha\beta)\langle Z_{a,b}\rangle.
\end{align}
We see that all four combinations $(\alpha,\beta)=(\pm1,\pm1)$ are consistent with genus one modular invariance -- but we should check genus two as well.  The modified decomposition relations there are
\begin{align}
\langle Z_{1,1,1,1}\rangle&\to\langle Z_{1,1,1,1}\rangle,\\
\langle Z_{1,1,1,-1}\rangle&\to 15\langle Z_{1,1,1,1}\rangle,\\
\langle Z_{1,1,1,i}\rangle&\to 8(1+\alpha)\langle Z_{1,1,1,a}\rangle,\\
\langle Z_{1,1,1,j}\rangle&\to 8(1+\beta)\langle Z_{1,1,1,b}\rangle,\\
\langle Z_{1,1,1,k}\rangle&\to 8(1+\alpha\beta)\langle Z_{1,1,1,c}\rangle,\\
\langle Z_{1,1,i,j}\rangle&\to 4(1+\alpha+\beta+\alpha\beta)\langle Z_{1,1,a,b}\rangle.
\end{align}
The trivial case $\alpha=\beta=1$ reproduces the results from section \ref{app:decomp}.  For the three non-trivial choices of quantum symmetry given by $(\alpha,\beta)=(1,-1)$, $(-1,1)$ or $(-1,-1)$, the $Q_8$ orbifold is equivalent to the orbifold by one of the three $\Z_2$ subgroups of $\Z_2\times\Z_2$, and this is consistent with both the genus one and genus two results.  So in this case, despite the fact that $\Gamma$ is non-abelian, the presence of the quantum symmetry causes it to be equivalent to an abelian orbifold and we do not need to worry about additional contributions to its decomposition.

\section{General Extensions and Quantum Symmetries}
\label{app:noncentral}

We repeat the analysis of section \ref{sec:classification}, now allowing $K$ to be a general finite group.  In particular $K$ need no longer be abelian or central in $\Gamma$.  Many of the arguments will carry through, though we will encounter some cases that present difficulties.

\subsection{A General Finite Extension}

We examine an extension of the finite group $G$ by another finite group $K$, given as before by
\be
1\to K\to \Gamma\to G\to 1.
\ee
In order to define the group law on $\Gamma$ we need a few ingredients.  Given a section $s(g): G\to \Gamma$, we can construct a map $\psi: G\to \text{Aut}(K)$ as
\be
\psi_g(k)=s(g)ks^{-1}(g).
\ee
When $K$ is abelian, $\psi$ gives it the structure of a $G$-module.

Additionally, we can construct an object
\be
c(g_1,g_2)=s(g_1)s(g_2)s^{-1}(g_1g_2).
\ee
We could quotient $K$ by its commutator subgroup to form its abelianization $K_{ab}$; the image of $c$ under this map belongs to a class in $H^2(G,K_{ab})$ which classifies the possible extensions of $G$ by $K$.  In a slight abuse of nomenclature we refer to both versions as the extension class.

Going forward we will abbreviate $s(g_i)=s_i$, $\psi_{g_i}(k)=\psi_i(k)$ and $c(g_i,g_j)=c_{i,j}$ in calculations.  Note that $\psi$ is a homomorphism in $K$ since
\be
\psi_{1}(k_2k_3)=s_1k_2k_3s^{-1}_1=s_1k_2s^{-1}_1s_1k_3s^{-1}_1=\psi_1(k_2)\psi_1(k_3).
\ee
We can see that $c$ satisfies a closure-like condition (which descends to genuine closure in $H^2(G,K_{ab})$) by comparing
\be
c_{1,2}c_{12,3}=s_1s_2s^{-1}_{12}s_{12}s_3s^{-1}_{123}=s_1s_2s_3s^{-1}_{123}
\ee
with
\be
\psi_1(c_{2,3})c_{1,23}=s_1s_2s_3s^{-1}_{23}s^{-1}_1s_1s_{23}s^{-1}_{123}=s_1s_2s_3s^{-1}_{123},
\ee
so our `closure' condition on $c$ is
\be
\label{c_closure}
dc=1=\psi_1(c_{2,3})c_{1,23}c^{-1}_{12,3}c^{-1}_{1,2}.
\ee
When $K$ is non-abelian, $\psi_g(k)$ will not be a homomorphism in $G$  -- we can see this from
\be
\label{psi_cond1}
\psi_{12}(k)=s_{12}ks^{-1}_{12}=s_{12}s^{-1}_2s^{-1}_1s_1s_2ks^{-1}_2s^{-1}_1s_1s_2s^{-1}_{12}=c^{-1}_{1,2}\psi_1(\psi_2(k))c_{1,2},
\ee
which also gives us
\be
\label{psi_cond2}
\psi_1(\psi_2(k))=c_{12}\psi_{12}(k)c^{-1}_{12}.
\ee
Now we can write down group multiplication in $\Gamma$.  It will be given by
\be
\label{group_multi}
(k_1,g_1)(k_2,g_2)=(k_1\psi_{g_1}(k_2)c(g_1,g_2),g_1g_2).
\ee
We can check that this is associative.  We have
\begin{multline}
((k_1,g_1)(k_2,g_2))(k_3,g_3)=(k_1\psi_1(k_2)c_{1,2},g_1g_2)(k_3,g_3)\\
=(k_1\psi_1(k_2)c_{1,2}\psi_{12}(k_3)c_{12,3},g_1g_2g_3)
\end{multline}
and
\begin{multline}
(k_1,g_1)((k_2,g_2)(k_3,g_3))=(k_1,g_1)(k_2\psi_2(k_3)c_{3,2},g_2g_3)\\
=(k_1\psi_1(k_2\psi_2(k_3)c_{2,3})c_{1,23},g_1g_2g_3).
\end{multline}
In order for these to be equal we must have
\be
k_1\psi_1(k_2)c_{1,2}\psi_{12}(k_3)c_{12,3}=k_1\psi_1(k_2\psi_2(k_3)c_{2,3})c_{1,23},
\ee
which holds if
\be
c_{1,2}\psi_{12}(k_3)c_{12,3}=\psi_1(\psi_2(k_3))\psi_1(c_{2,3})c_{1,23}.
\ee
Applying (\ref{psi_cond1}), this reduces to the closure condition (\ref{c_closure}) which we have already shown, so our group multiplication is associative.

Finally we check inverses.  We claim that the inverse of $(k,g)$ is given by
\be
\label{inverse}
(k,g)^{-1}=(c^{-1}(g^{-1},g)\psi_{g^{-1}}(k^{-1}),g^{-1}).
\ee
From the left this is straightforward to check:
\begin{multline}
(c^{-1}(g^{-1},g)\psi_{g^{-1}}(k^{-1}),g^{-1})(k,g)\\
=(c^{-1}(g^{-1},g)\psi^{-1}_{g^{-1}}(k)\psi_{g^{-1}}(k)c(g^{-1},g),g^{-1}g)=(1,1).
\end{multline}
To see that it is also a right inverse, we first note that the closure condition (\ref{c_closure}) on $c$ implies that
\be
\psi_g(c^{-1}(g^{-1},g))=c^{-1}(g,g^{-1}).
\ee
Now we calculate
\begin{multline}
(k,g)(c^{-1}(g^{-1},g)\psi_{g^{-1}}(k^{-1}),g^{-1})=(k\psi_g(c^{-1}(g^{-1},g))\psi_g(\psi_{g^{-1}}(k^{-1}))c(g,g^{-1}),1)\\
=(kc^{-1}(g,g^{-1})\psi_g(\psi_{g^{-1}}(k^{-1}))c(g,g^{-1}),1)=(k\psi_{gg^{-1}}(k^{-1}))=(kk^{-1},1)=(1,1),
\end{multline}
so this inverse works from both sides.

\subsection{Including Quantum Symmetries}

In this section we work with the object $B(k,g)$ which is a homomorphism in $K$.  Since $K$ is not central in $\Gamma$ we will have an action of $G$ on $\phi\in H^1(K,U(1))$ given by
\be
g\cdot \phi(k) = \phi(\psi_{g^{-1}}(k))
\ee
which satisfies
\be
g_1\cdot g_2\cdot \phi(k) = \phi(\psi_{g_2^{-1}}(\psi_{g_1^{-1}}(k))).
\ee
Since we are working in a homomorphism, (\ref{psi_cond1}) allows us to swap $\psi_1(\psi_2)$ with $\psi_{12}$ so we have
\be
g_1\cdot g_2\cdot \phi(k)=\phi(\psi_{(g_1g_2)^{-1}}(k))=(g_1g_2)\cdot\phi(k).
\ee

Again we begin under the assumption that $K$ acts non-trivially on the parent theory, so to guarantee modular invariance $B$ should assign the same phase to each partial trace in a given $\Gamma$ orbit.  Consider the partial trace $Z_{(k_1,g_1),(k_2,g_2)}$.  The possible modifications to the action of $K$ are given by a slight modification of (\ref{kgkg}) as
\be
\label{kgkg2a}
\frac{B(\psi_{g_1}(k_2),g_1)}{B(\psi_{g_2}(k_1),g_2)}Z_{(k_1,g_1),(k_2,g_2)}
\ee
where, as before, $B$ is a homomorphism in its $K$ argument and is normalized such that $B(1,g)=B(k,1)=1$.  Consider the modular transformation
\begin{multline}
Z_{(k_1,g_1),(k_2,g_2)}(\tau-1)=Z_{(k_1,g_1),(k_1,g_1)(k_2,g_2)}(\tau)\\
=Z_{(k_1,g_1),(k_1\psi_1(k_2)c_{1,2},g_1g_2)}(\tau).
\end{multline}
Performing this transformation on (\ref{kgkg2a}) yields
\be
\frac{B(\psi_1(k_2),g_1)}{B(\psi_2(k_1),g_2)}Z_{(k_1,g_1),(k_1\psi_1(k_2)c_{1,2},g_1g_2)}.
\ee
Swapping the order of the modular transformation and phase assignments yields
\be
\frac{B(\psi_1(k_1\psi_1(k_2)c_{1,2}),g_1)}{B(\psi_{12}(k_1),g_1g_2)}Z_{(k_1,g_1),(k_1\psi_1(k_2)c_{1,2},g_1g_2)},
\ee
so consistency requires that
\be
\label{epsilon_consistencya}
\frac{B(\psi_1(k_2),g_1)}{B(\psi_2(k_1),g_2)}=\frac{B(\psi_1(k_1\psi_1(k_2)c_{1,2}),g_1)}{B(\psi_{12}(k_1),g_1g_2)}.
\ee
The commutation conditions on the two elements of $\Gamma$ are
\be
k_1\psi_1(k_2)c_{1,2} = k_2\psi_2(k_1)c_{2,1}\hspace{.5cm}\text{and}\hspace{.5cm}g_1g_2=g_2g_1,
\ee
and we can use this commutation relation to rewrite the numerator on the rhs of (\ref{epsilon_consistencya}) to get
\be
\frac{B(\psi_1(k_2),g_1)}{B(\psi_2(k_1),g_2)}=\frac{B(\psi_1(k_2\psi_2(k_1)c_{1,2}),g_1)}{B(\psi_{12}(k_1),g_1g_2)}.
\ee
Using the fact that $B$ is a homomorphism in its $K$ argument we can both split up the numerator of the rhs and replace $\psi_1(\psi_2)$ with $\psi_{12}$, giving
\be
\label{crossedhomomorphism}
B(\psi_{12}(k_1),g_1g_2)=B(\psi_{12}(k_1),g_1)B(\psi_2(k_1),g_2)B(\psi_1(c_{1,2}),g_1).
\ee
Once again we find an obstruction to modular invariance when $B(\psi_{g_1}(c(g_2,g_3)),g_1)$ is not trivial.  The condition that $B$ is a homomorphism in $G$ has been replaced with
\be
\label{crossed_homomorphism}
B(k,g_1g_2)=B(k,g_1)B(\psi_{g_1^{-1}}(k),g_2)=B(k,g_1)\text{ }g_1\cdot B(k,g_2),
\ee
which is the condition for being a crossed homomorphism in $G$, i.e. such a quantum symmetry lives in $Z^1(G,H^1(K,U(1)))$: one-cochains in $G$ valued in $H^1(K,U(1))$ (with non-trivial action on the coefficients).

What do exact elements of $Z^1(G,H^1(K,U(1)))$ look like?  They take the form
\be
B(k,g)=\frac{g\cdot \varphi(k)}{\varphi(k)}=\frac{\varphi(\psi_{g^{-1}}(k))}{\varphi(k)}
\ee
for $\varphi\in H^1(K,U(1))$.  When the extension is split it is easy to see that these do not contribute.
If we make such a choice for $B$, a generic partial trace $Z_{(k_1,g_1),(k_2,g_2)}$ receives the phase
\be
\label{exact_arg}
\frac{B(\psi_{g_1}(k_2),g_1)}{B(\psi_{g_2}(k_1),g_2)}=\frac{\varphi(k_2)\varphi(\psi_2(k_1))}{\varphi(k_1)\varphi(\psi_1(k_2))}=\frac{\varphi(k_2\psi_2(k_1))}{\varphi(k_1\psi_1(k_2))},
\ee
which we see precisely vanishes for $(k_1,g_1)$ and $(k_2,g_2)$ commuting.  Therefore this choice of quantum symmetry, while non-trivial, does not change the coefficients of decomposition, and we can mod out by $B$ which are exact.  This means our choice of $B(k,g)$ depends solely on its class in $H^1(G,H^1(K,U(1)))$.  When the extension is not split, (\ref{exact_arg}) fails to be trivial by
\be
\label{exact_obstruction}
\varphi(c(g_2,g_1)c^{-1}(g_1,g_2)).
\ee
This would seem to disrupt the conclusion that quantum symmetries for such extensions are classified by $H^1(G,H^1(K,U(1)))$ since we do not seem to be able to disregard exact elements.  Our suspicion is that the above analysis is incomplete, and that $H^1(G,H^1(K,U(1)))$ should always classify quantum symmetries -- this may be because our assignment of phases (\ref{kgkg2a}) needs to be further modified (that expression was assumed rather than derived).  As far as our examples in section \ref{sec:examples} are concerned, the above issue can only arise in extensions which are neither central nor split.  We present one such example with $\Gamma=Q_8$, the group of unit quaternions, in section \ref{sec:quaternion_extension}.  In that example the extension class is symmetric, so (\ref{exact_obstruction}) presents no issue.

Finally, in passing, we note that (\ref{dt_obstruction}) holds with the expected modification -- the obstruction to $B(\psi_1(k_2),g_1)$ defining a two-cocycle on $\Gamma$ is $B(\psi_1(c_{2,3}),g_1)$.  So once again quantum symmetries that are equivalent to discrete torsion are compatible with modular invariance in $\Gamma$.


\end{document}